\documentclass[aps,prl,reprint,amssymb,amsfonts,groupedaddress,longbibliography]{revtex4-1}

\usepackage{graphicx}
\usepackage{dcolumn}
\usepackage{bm}

\usepackage[utf8]{inputenc}
\usepackage{mathptmx}

\usepackage{upgreek} 
\usepackage{xcolor} 

\begin{document}

\title[Absorber Coupling Design]{Impact of the absorber coupling design for Transition Edge Sensor X-ray Calorimeters}

\author{M. de Wit}
 \email{M.de.Wit@sron.nl}
\author{L. Gottardi}
\author{E. Taralli}
\author{K. Nagayoshi}
\author{M.L. Ridder} 
\author{H. Akamatsu}
\author{M.P. Bruijn}
\author{R.W.M. Hoogeveen}
\author{J. van der Kuur}
\author{K. Ravensberg}
\author{D. Vaccaro} 
\author{J-R. Gao}
 \altaffiliation{Faculty of Applied Science, Delft University of Technology, Lorentzweg 1, 2628 CJ Delft, The Netherlands}
\author{J-W.A. den Herder}
 \affiliation{NWO-I/SRON Netherlands Institute for Space Research, Niels Bohrweg 4, 2333 CA Leiden, The Netherlands}

\date{\today}

\begin{abstract}
Transition Edge Sensors (TESs) are the selected technology for future spaceborne X-ray observatories, such as Athena, Lynx, and HUBS. These missions demand thousands of pixels to be operated simultaneously with high energy-resolving power. To reach these demanding requirements, every aspect of the TES design has to be optimized. Here we present the experimental results of tests on different devices where the coupling between the x-ray absorber and the TES sensor is varied. In particular, we look at the effects of the diameter of the coupling stems and the distance between the stems and the TES bilayer. Based on measurements of the AC complex impedance and noise, we observe a reduction in the excess noise as the spacing between the absorber stem and the bilayer is decreased. We identify the origin of this excess noise to be internal thermal fluctuation noise between the absorber stem and the bilayer. Additionally, we see an impact of the coupling on the superconducting transition in the appearance of kinks. Our observations show that these unwanted structures in the transition shape can be avoided with careful design of the coupling geometry. Also the stem diameter appears to have a significant impact on the smoothness of the TES transition. This observation is still poorly understood, but is of great importance for both AC and DC biased TESs.
\end{abstract}

\maketitle

\section{Introduction} \label{sec:intro}

Transition Edge Sensors (TESs) are very sensitive detectors of energy that utilize the sharpness of a superconductor normal-to-superconducting transition\cite{Irwin2005, Ullom2015, Gottardi2021a}. TESs are the selected technology in a number of either planned or proposed space-borne observatories, such as Athena \cite{Barret2020a}, Lynx \cite{Gaskin2019}, and HUBS \cite{Cui2020}. These missions aim for thousands of pixels to be operated simultaneously with high energy resolving power. The limited resources available for a satellite-based instrument, such as electrical power, cooling power, and allowable weight, require the use of sophisticated multiplexing schemes, such as Time Division Multiplexing (TDM) \cite{Durkin2019}, Frequency Division Multiplexing (FDM) \cite{Akamatsu2020}, and others \cite{Morgan2016,Bennett2019,Nakashima2020}. In this paper, we focus on TES based X-ray micro-calorimeters developed for the Athena X-ray Integral Field Unit (X-IFU) instrument \cite{Pajot2018}. X-IFU will consist of over 3000 multiplexed pixels with a planned spectral resolution of 2.5$~$eV for 7$~$keV X-rays.

In order to achieve the demanding requirements set by the X-IFU instrument, every aspect of the TES pixel design has to be optimized. Over the years much work has been done towards this goal. Research topics included the impact of the thermal conductance on the TES transition width \cite{Hays-Wehle2016, Morgan2017, Zhang2019}, the impact of the normal resistance by changing the TES size and aspect ratio \cite{Gottardi2017, Sakai2018, Morgan2019, DeWit2020}, the noise-mitigating properties of normal-metal features \cite{Ullom2004, Sadleir2011, Miniussi2018, Wakeham2018}, and how the energy deposited in the absorbed is transferred to the bilayer \cite{Kilbourne2007a, Smith2012, Smith2014}. In this paper, we expand on this last point by studying how the detailed design of the coupling between the X-ray absorber and the TES bilayer affects the properties of the TES micro-calorimeters. The basic layout of the TESs are explained in Sec. \ref{sec:setup} of this paper.

Previous work has shown that the design of the coupling geometry could have a big impact on the detector properties. It was demonstrated that the stems introduce effects of non-equilibrium superconductivity via the proximity effect, affecting the critical temperature and critical current, and influence the small-signal transition parameters $\alpha = \frac{T}{R} \frac{\partial R}{\partial T}$ and $\beta = \frac{I}{R} \frac{\partial R}{\partial I}$, and the detector energy resolution and time constants \cite{Bandler2008, Sadleir2011, Smith2013, Smith2014}. Note that these effects were reported for stems with a very large footprint on the bilayer, while in this work, we study the impact of much smaller stems, which will allow us to pick up on more subtle differences than those reported before. We mainly investigate the effect of the diameter of the two stems that connects the absorber to the bilayer, and the spacing between these two. We measured the complex impedance and noise, which are used to estimate the internal thermal conductance in the TES and how this is influenced by the geometry. We compare the small-signal limit energy resolution, estimated from the measured noise and transition shape, with the measured Noise Equivalent Power (NEP), and find a good qualitative agreement between the two.

\section{Detectors and Measurement Setup} \label{sec:setup}

\begin{figure*}[ht!]
\centering
\includegraphics[width=0.95\linewidth]{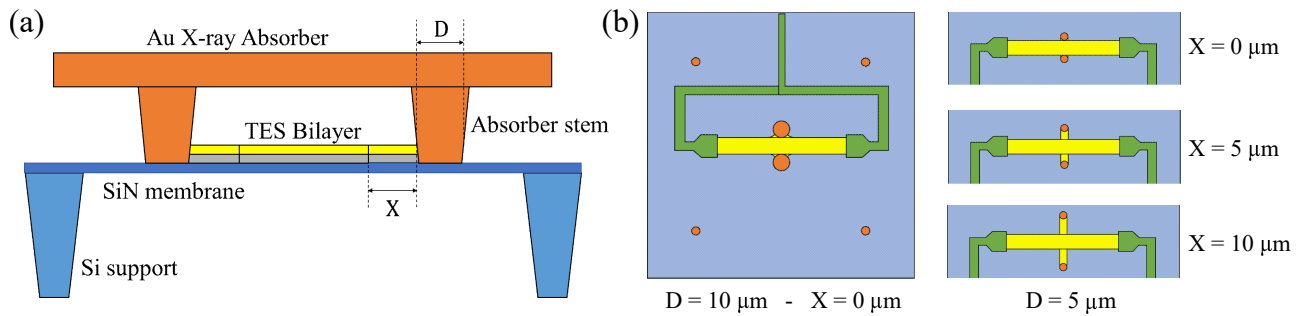}
\caption{(a) Schematic of the layout of the TESs. We vary only two aspects of the design: the diameter of the absorber stem $D$, and the spacing between the bilayer and the stem $X$. (b) Overhead view of the 4 different designs studied: A single design with $D = 10~\upmu$m and $X = 0~\upmu$m, and three designs with $D = 5~\upmu$m and varying spacing, $X = 0$, $5$, and $10~\upmu$m. The green structures are the niobium leads.}
\label{figure:TES_Layout}
\end{figure*}

We have fabricated an $8 \times 8$ TES array with four different geometries for the absorber-bilayer coupling. The basic layout of the detectors is shown schematically in Fig. \ref{figure:TES_Layout}(a). The TESs are based on a bilayer made of 35$~$nm Ti and 200$~$nm Au, with a sheet resistance of 26~m$\Omega$/$\square$. The bilayer is patterned to be an $80 \times 10~\upmu$m$^2$ rectangle with a normal resistance of approximately 200~m$\Omega$. Niobium leads are used to apply a voltage bias to the devices. The bilayers are grown on a 0.5$~\upmu$m thick silicon nitride membrane. The TES is connected to an Au X-ray absorber via two central absorber stems with a height of 3.5~$\upmu$m, one at each edge of the bilayer. Four additional stems (not shown in Fig. \ref{figure:TES_Layout}(a)) with diameter 5~$\upmu$m are used to support the absorber on the membrane. The absorber has a thickness of 2.35~$\upmu$m and a heat capacity of 0.85~pJ/K at 90~mK. Further details about the TESs and the fabrication procedure are given by \citeauthor{Nagayoshi2019}\cite{Nagayoshi2019}.

To study how the detailed coupling between the absorber and the bilayer affects the detector properties, we vary the geometry of the two absorber stems, as schematically illustrated in Fig. \ref{figure:TES_Layout}(b). We look at the possible impact of the diameter of the two central stems, by comparing two designs with diameters $D = 10~\upmu$m and $D = 5~\upmu$m, both fabricated such that the edge of the bilayer is aligned with the edge of the outer perimeter of the stem. Additionally, we look at whether the precise position of the stems is of importance for the final detector performance, by comparing three designs with the same diameter $D = 5~\upmu$m but different spacing between the edge of the bilayer and the edge of the stems, $X = 0$, $5$, and $10~\upmu$m. The arm connecting the stem and the bilayer is an integral part of the bilayer, deposited in the same step of the fabrication. All other properties of the devices are identical to allow for direct comparison.

We characterize the devices using an FDM setup, with which we bias a single pixel at a time within the superconducting transition using an alternating current (AC). The resistance of the TES at a particular bias point is generally specified with respect to the normal resistance $R_n$. Each of the investigated pixels is connected to a high-Q LC resonator \cite{Bruijn2012} with a bias frequency between 1-5 MHz. Further details about the FDM setup are given elsewhere \cite{Akamatsu2016a, DeWit2020}. Most of the results presented in the remainder of this paper do not depend on the bias frequency, and are valid for both AC and DC biased pixels. Therefore, we focus on presenting results from pixels measured at low bias frequencies ($<2~$MHz), as these results are typically more straightforward to interpret and compare to the literature.

The setup is mounted at the mixing-chamber of a cryogen-free dilution refrigerator, and is stabilized at a temperature of 50$~$mK. External magnetic fields are shielded using a room-temperature $\mu$-metal shield and a lead and cryoperm shield at the cold stage. Both the setup and the cryoperm shield are suspended from a kevlar vibration isolation stage for mechanical decoupling \cite{Gottardi2019a}. A more extensive overview of the setup is given elsewhere \cite{DeWit2020}.

For all devices measurements of the current-voltage characteristic (IV-curve) at different bath temperatures are used to determine the critical temperature and thermal conductance of the detectors. The averaged thermal conductance for all devices is 46.5~pW/K with a standard deviation of 4.6~pW/K at a critical temperature of $81.8 \pm 0.2~$mK. Given the fact that the thermal conductance of this type of devices is generally proportional to the phonon-emitting perimeter of the TES \cite{Hoevers2005, Hays-Wehle2016, DeWit2020}, the thermal conductance is expected to increase for larger $X$, with a difference of approximately 7 pW/K between the $X = 0$ and $10~\upmu$m devices. This predicted difference has not been resolved in the measured data. We have observed no effect of the stem geometry on the current distribution, nor in the temperature dependence of the critical current.

\section{Impact of Stem-Bilayer Distance} \label{sec:distance}

\begin{figure}
\centering
\includegraphics[width=1.0\columnwidth]{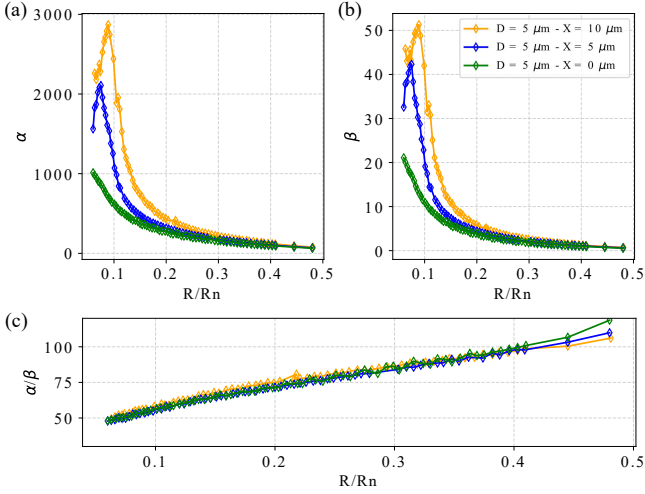}
\caption{Results of complex impedance and noise measurements for low bias frequency pixels with different stem-bilayer separations. $\alpha$ (a) and $\beta$ (b) extracted from complex impedance measurements for the relevant part of the normal to superconducting transition. (c) Ratio of $\alpha$/$\beta$ computed for the same bias points as those shown in (a) and (b).}
\label{figure:alpha_beta}
\end{figure}

We fully characterize the properties of the transition by measuring the complex impedance of the TES \cite{Lindeman2004}. For various points in the transition, a small modulation is added to the bias current with a frequency ranging from $10~$Hz$~-~50~$kHz, where we record the TES response for each frequency. This data is combined with knowledge of the heat capacity, which is dominated by the X-ray absorber, and thermal conductance to extract the dimensionless derivatives of the resistance with respect to the temperature $\alpha$ and to the current $\beta$. We focus on these parameters, as they describe the shape of the superconducting transition and can be used to obtain, among other things, the bandwidth and sensitivity of the detectors. Details of this measurement and analysis under AC bias are given elsewhere \cite{Taralli2019a, Gottardi2021}. Representative data for $\alpha$ and $\beta$ for low bias frequency devices with varying length of the arm between the absorber and the bilayer is shown in Fig. \ref{figure:alpha_beta}.

All the devices show a similar ratio between $\alpha / \beta$, visible in Fig. \ref{figure:alpha_beta}(c). However, there are clear differences between the devices low in the resistive transition, as visible in Figs. \ref{figure:alpha_beta}(a) and (b). The $X = 5~\upmu$m and $X = 10~\upmu$m devices show high peaks in both $\alpha$ and $\beta$, signs of sharp features or kinks in the superconducting transition. No kinks are observed for the $X = 0~\upmu$m devices for all measured bias frequencies. Kinks are regularly observed in devices with normal metal stripes on top of the TES \cite{Smith2015, Wakeham2018}. Here we show that also the normal metal of the absorber stems might play a role in the precise location of these kinks in the transition. For the shortest arms with $X = 0~\upmu$m, the kinks are absent or might have shifted to lower resistance values outside of the region of interest. The removal of kinks from the relevant region of the parameter space is of utmost importance for achieving uniform performance in large arrays of TESs.

\begin{figure*}[t!]
\centering
\includegraphics[width=1.0\linewidth]{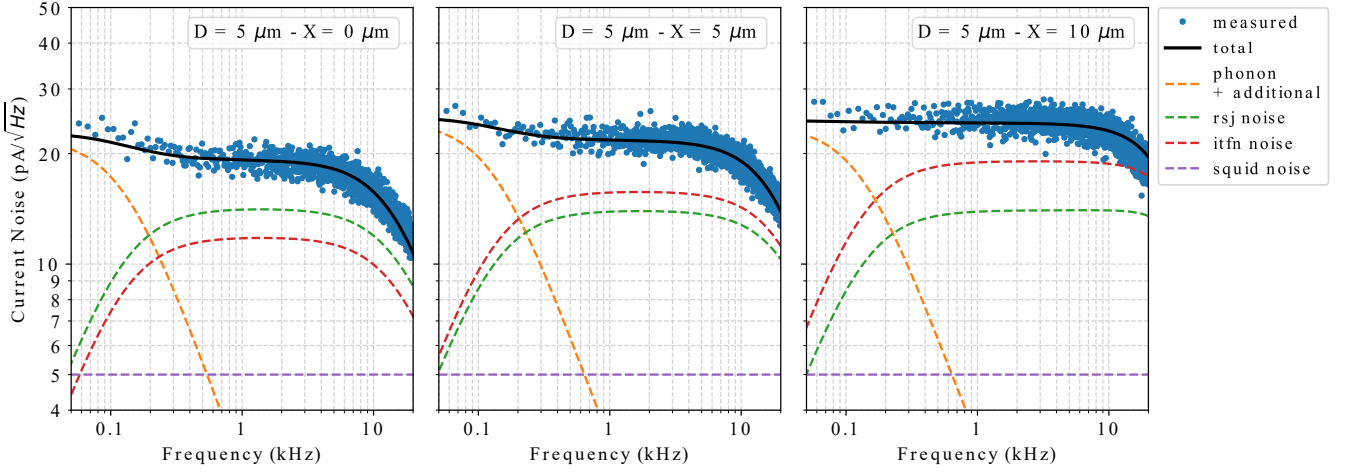}
\caption{Example noise spectra measured for the different $D = 0~\upmu$m pixels, measured at $R \sim 0.15~R_n$. Dashed lines represent the calculated noise models: squid noise (purple), phonon and additional low frequency noise (orange), voltage noise from RSJ model (green), and ITFN (red). The total of all noise sources is given by the solid black line.}
\label{figure:spectra}
\end{figure*}

Apart from the complex impedance, for each bias point we have measured the current noise spectrum. Example noise spectra for the three $D = 5~\upmu$m pixels measured at $R = 0.15~R_n$ are shown by the blue points in Fig. \ref{figure:spectra}. The roll-off visible at high frequencies results from the bandwidth of the electrical circuit.

\begin{figure}
\centering
\includegraphics[width=0.95\linewidth]{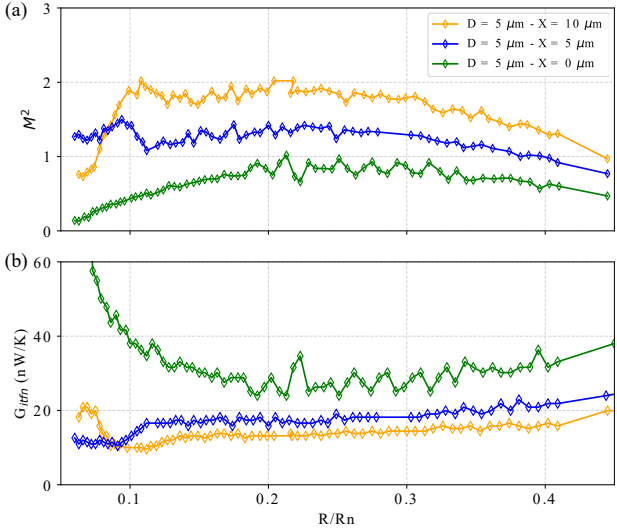}
\caption{(a) Excess noise factor defined with respect to expected noise in the RSJ model $S_{V,RSJ}$ for the same bias points show in Fig. \ref{figure:alpha_beta}. (b) $G_{\rm{itfn}}$ required to minimize the excess noise factor for each bias point.}
\label{figure:M2_Gitfn}
\end{figure}

We can compare the measured noise spectra with the calculated contributions from various noise sources. The principal sources of noise include the white readout noise of the SQUID, low frequency phonon noise coming from thermal fluctuations between the TES and the bath (and other low frequency noise sources such as vibrations and stray photons, all summed under the name additional low frequency noise), and Johnson noise from the TES resistance. It has recently been shown that for this type of devices the Johnson noise is best described by the Resistively-Shunted-Junction (RSJ) model \cite{Kozorezov2012, Gottardi2021, Wessels2021}, given by:
\begin{equation}
S_{V,RSJ} = 4k_BTR \left( 1 + \frac{5}{2}\beta + \frac{3}{2}\beta^2 \right),
\end{equation}
with $k_B$ being the Boltzmann constant. These main sources of noise are shown as the various dashed lines in Fig. \ref{figure:spectra}. However, the sum of these noise sources is clearly insufficient to fully explained the measured noise. Therefore, we express the factor of difference between the measured noise and the predicted noise using the parameter $M^2$, defined by: 
\begin{equation}
M^2 = \left( S_{V,data} - S_{V,RSJ} \right) / S_{V,data}
\end{equation}
Fig. \ref{figure:M2_Gitfn}(a) shows $M^2$ for various bias points in the transition for the different designs, showing a clear rise in $M^2$ as $X$ increases.

In order to explain the additional noise in the kHz range, we interpret it as resulting from internal thermal fluctuation noise (ITFN) within the detector itself \cite{Wakeham2019, Gottardi2021}. ITFN occurs when there are distributed heat capacities inside the detector connected by internal thermal impedances. Such a complex thermal structure results in thermal fluctuations between the different heat capacities. To confirm this hypothesis we add another noise contribution coming from the ITFN, given by \mbox{$S_{P,\rm{itfn}} = 4k_B T^2 G_{\rm{itfn}}$}, where $G_{\rm{itfn}}$ is a thermal conductance located within the TES \cite{Palosaari2012, Maasilta2012a}. This contribution is shown as the red dashed lines in Fig. \ref{figure:spectra}. Note that the shape of the ITFN is similar to the RSJ Johnson noise due to the conversion of power to current noise via the power-to-current responsivity and the TES circuit inductance. $G_{\rm{itfn}}$ is treated as a free parameter to be varied such that $M^2$ is minimized. The resulting $G_{\rm{itfn}}$ is shown in Fig. \ref{figure:M2_Gitfn}(b). As expected, the internal thermal conductance decreases as $X$ increases. At $R/R_n \sim~20$\%, the extracted $G_{iftn}$ increases from $\sim14~$nW/K for $X = 10~\upmu$m to $\sim18~$nW/K for $X = 5~\upmu$m and finally to $\sim27~$nW/K for $X = 0~\upmu$m.

\begin{figure}
\centering
\includegraphics[width=1.0\linewidth]{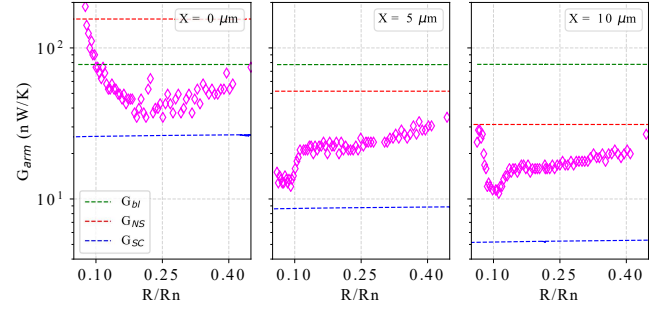}
\caption{Estimated value of $G_{arm}$, assuming a constant bilayer thermal conductance given by the Wiedemann-Franz law (green dashed line). Blue dashed line shows the arm thermal conductance when it remains superconducting (ignoring the temperature dependence of the gap). The red dashed line shows the arm thermal conductance when it is in the normal state (and thus follows the Wiedemann-Franz law). We take the effective length of the arm equal to $X + D/2$.}
\label{figure:Garm}
\end{figure}

The internal thermal conductance is believed to be a combination of the internal thermal conductance of the bilayer $G_{bl}$, and the limited thermal conductance of the arm between the absorber stem and the bilayer, $G_{arm}$. Using the simplest possible thermal model in which the different thermal components are connected in parallel, $G_{arm}$ can be estimated from the measured $G_{\rm{itfn}}$ using:
\begin{equation}
G_{arm} = \left(\frac{1}{G_{\rm{itfn}}} - \frac{1}{G_{bl}}\right)^{-1}
\end{equation}
The internal thermal conductance of the bilayer is estimated by using the Wiedemann-Franz law, $G_{bl} = L_0 T / R_\square$, with $L_0 = 2.45\cdot 10^{-8}~\Omega W K^{-2}$ and $R_\square$ the sheet resistance of the bilayer \cite{Gottardi2021}. For the $80 \times 10 \upmu$m$^2$ devices, $G_{bl} = 77$ nW/K. The estimated $G_{arm}$ is shown as the magenta points in Fig. \ref{figure:Garm}.

To understand the origin of the limited thermal conductance of the arm, we estimate it in two limiting regimes. The upper limit is given by the normal state thermal conductance of the arm $G_{NS}$, which again is estimated using the Wiedemann-Franz law.  In this case the resistance of the arm is calculated using an effective length $X + D/2$, the full length of the arm including the end part covered by the absorber stem. This upper limit is shown in Fig. \ref{figure:Garm} as the red dashed lines. The lower limit is obtained by assuming that the arm remains superconducting. As cooper pairs cannot transport heat the thermal conductance is purely governed by the fraction of depaired electrons, which is dictated by the temperature dependent superconducting gap $\Delta(T)$ and the thermal excitation energy $k_B T$. Thus, the thermal conductance can be estimated using:
\begin{equation}
G_{SC} = G_{NS} ~ \exp{\left( - \frac{\Delta(T)}{k_B T} \right)}
\end{equation}
In order to evaluate this equation knowledge about the temperature dependence of the superconducting gap is required. For a conventional s-wave superconductor at $T \approx T_C$, this dependence is given by
\begin{equation}
\Delta(T) \approx 1.74 \Delta(0) \sqrt{1 - \frac{T}{T_C}},
\end{equation}
with $\Delta(0) = 1.764~k_B T_C$ the gap at T = 0 \cite{Tinkham1996}. However, this is not valid in the presence of the proximity effect, which significantly alters the size of the gap close to the $T_C$ of the bilayer. Therefore, we simplify the problem and derive a lower limit for the thermal conductance by neglecting the temperature dependence of the gap and using $\Delta(T) = \Delta(0)$. The resulting $G_{SC}$ is shown as the blue dashed lines in Fig. \ref{figure:Garm}. Indeed the values for $G_{arm}$ extracted from the experimental data fall between these two boundaries, supporting the interpretation that the arm partially remains superconducting with a $\Delta(T)$ that is suppressed compared to the ideal $T = 0$ value, but remains finite.

The previous analysis clearly indicates that a small but significant reduction of the excess noise can be achieved by keeping the stem as close to the bilayer as possible, or perhaps by even placing the stems on top of the bilayer\cite{Wakeham2020}. Note that after reducing the ITFN, the performance of this type of devices under these conditions will remain limited by the noise resulting from the weak-link effect as described by the RSJ model.

\section{Impact of Stem-Bilayer Diameter} \label{sec:diameter}

\begin{figure}
\centering
\includegraphics[width=0.9\columnwidth]{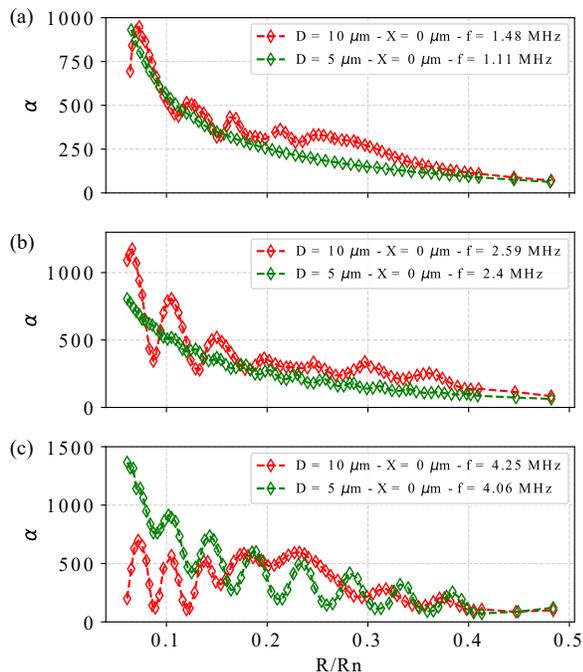}
\caption{Measured $\alpha$ versus $R/R_n$ for pixels with different stem diameters. We show three sets of pixels at different bias frequencies: (a) 1.1 - 1.5 MHz, (b) 2.4 - 2.6 MHz, and (c) 4.1 - 4.3 MHz.}
\label{figure:D_alpha}
\end{figure}

\begin{figure}
\centering
\includegraphics[width=0.95\columnwidth]{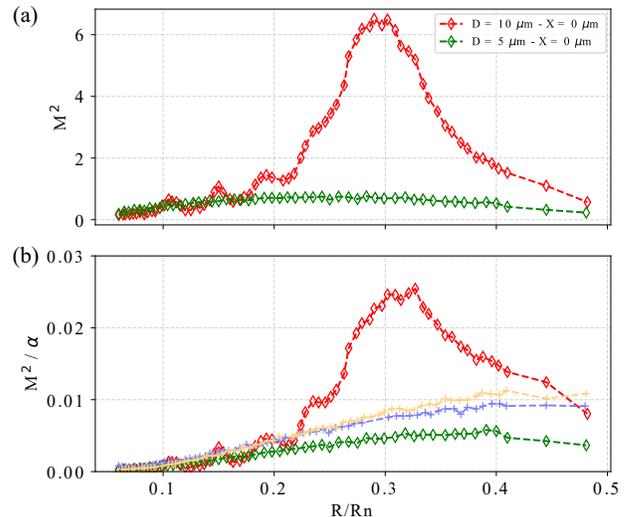}
\caption{(a) Excess noise $M^2$ and (b) $M^2$ / $\alpha$ for low freq pixels with different stem diameters. In (a), data are shown for the $X = 0,~D = 10~\upmu$m (red) and $X = 0,~D = 5~\upmu$m (green) devices. In (b), also the data for the $X = 5~\upmu$m (blue) and $X = 10~\upmu$m (yellow) devices are shown as a reference.}
\label{figure:Excess_Noise}
\end{figure}

The main effects of the stem diameter are visible in the measured $\alpha$, shown in Fig. \ref{figure:D_alpha}, and excess noise, shown in Fig. \ref{figure:Excess_Noise}. In these figures, we focus on the designs with $X = 0~\upmu$m. Three different sets of pixels are selected, measured at (a) low, (b) mid, and (c) high bias frequencies, the precise values of which are indicated in the legend of Fig. \ref{figure:D_alpha}.

Changing the stem diameter from $D = 10$ to $D = 5~\upmu$m does not change the overall trend of $\alpha$. However, there is a striking difference between how $\alpha$, and consequently $\beta$ and $M^2$, changes throughout the transition. The $D = 10~\upmu$m devices show large oscillations of $\alpha$ for different bias points. These oscillations are typically attributed to the weak-link effect in these detectors \cite{Gottardi2018a, DeWit2020} resulting from the interaction between the high $T_C$ Nb leads and the low $T_C$ bilayer. However, the amplitude of the oscillations is much smaller for all $D = 5~\upmu$m devices, except for those measured at the highest bias frequencies as visible in Fig. \ref{figure:D_alpha}(c). In order to confirm this observation and to understand the underlying mechanism for why the smaller stem diameter improves the smoothness of the superconducting transition, more devices with different stem diameters should be studied.

Another interesting feature of the smaller stem diameter is visible in the excess noise in Fig. \ref{figure:Excess_Noise}(a). For all the devices studied in this work with $D = 10~\upmu$m, a peak in the excess noise is visible between $R = 0.2 ~-~ 0.5 R_n$. This peak is never observed for any of the devices with $D = 5~\upmu$m. These devices do show a bump in the transition around $R \approx 0.1 R_n$, visible in Fig. \ref{figure:alpha_beta}, but in those cases the increase in alpha is correlated with an increase in $M^2$. This is not the case for the $D = 10~\upmu$m devices, a fact that becomes clear when looking at Fig. \ref{figure:Excess_Noise}(b), in which we plot $M^2/\alpha$. The stark difference in the ratio of $M^2$ and $\alpha$ between the $D = 10~\upmu$m data (red points) and the other data signifies the difference in nature of the bump in the excess noise.

The exact origin for the excess noise bumps is unclear, but we believe that it's related to a scattering or interference effect of the quasi-particles interacting with the normal metal stems. A new set of devices is currently under fabrication to study these effects in more detail.

\section{NEP and energy resolution} \label{sec:NEP}

\begin{figure}
\centering
\includegraphics[width=0.95\columnwidth]{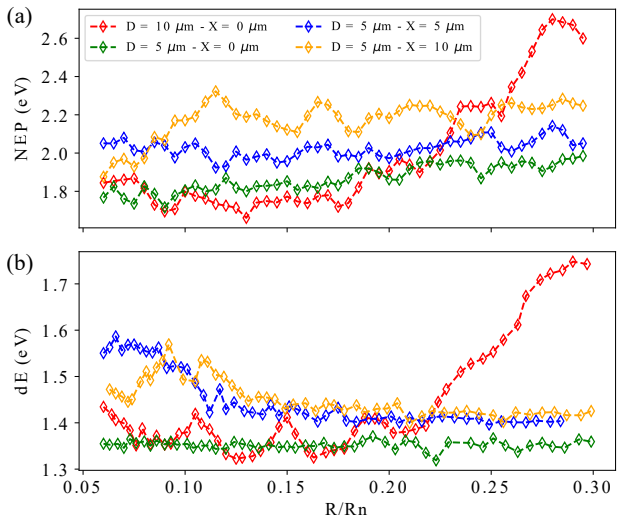}
\caption{(a) Measured NEP and (b) small-signal limit energy resolution calculated using Eq. \ref{eq:dE}. The NEP is measured at the optimal bias frequency for each value of $R/R_n$.}
\label{figure:NEP_dE}
\end{figure}

Arguably the most important figure of merit for a micro-calorimeter is the energy resolution, i.e. the capacity to resolve different photon energies. In this section, we demonstrate the effect of the absorber coupling geometry in two ways. First, we evaluate the integrated noise-equivalent power (NEP) at 5.9$~$keV for bias points between $R = 0.05 ~-~ 0.30 R_n$. The NEP is generally a good predictor for the achievable energy resolution in the large-signal limit. Secondly, we can use $\alpha$, $\beta$, and $M^2$, as shown in Secs. \ref{sec:distance} and \ref{sec:diameter}, to calculate the expected energy resolution in the small-signal limit for each bias point, using:
\begin{equation} \label{eq:dE}
\Delta E = 2\sqrt{2 \ln{2}} \sqrt{ 4 k_B T^2 \frac{C}{\alpha} \sqrt{ \left( 1 + \frac{5}{2}\beta + \frac{3}{2}\beta^2 \right) \left( 1+M^2 \right)} }
\end{equation}
The results for the NEP and expected energy resolution are shown in Fig. \ref{figure:NEP_dE} for four different pixels, one of each design, measured at low bias frequencies. In absolute terms, a relatively large difference is observed between the NEP and estimated $\Delta E$. This is most likely the result from the fact that our detectors are working outside of the small-signal limit for 5.9 keV X-rays. Note that the increase in NEP and $\Delta E$ seen in the $X = 0~\upmu$m device at higher $R/R_n$ is caused by the bump in the excess noise shown in Fig. \ref{figure:Excess_Noise}.

The NEP and $\Delta E$ qualitatively give the same results, and confirm the trends previously identified. The best energy resolutions are expected for the devices with $X = 0~\upmu$m. As the length of the arm between the absorber and the bilayer increases, the energy resolution becomes worse. This observation holds over the full bias frequency range. The increase in the energy resolution is about $0.2 ~-~ 0.4$~eV, and clearly indicates that maximizing the thermal conductance between the absorber and the bilayer is of great importance to optimize the resolution. The effect of the stem diameter is much less pronounced, with only a very small reduction of less than 0.1~eV for both the NEP and $\Delta E$. This is not completely unexpected. If we assume that the gold in the absorber stem has a electrical resistivity $\rho_{Au} = 1 $n$\Omega$m, similar to previously reported values \cite{Wakeham2019}, the Wiedemann-Franz law can be used to estimate a thermal conductance of the stem (with height 3.5 $\upmu$m). This leads to $G_{stem} = 11~\upmu$W/K and $45~\upmu$W/K for the $5~\upmu$m and $10~\upmu$m diameter stems, respectively. For both diameters, $G_{stem}$ is much higher that the estimated thermal conductance of the arm, and therefore does not play a significant role in the noise.

\section{Conclusions}

We have studied the impact of the detailed geometry for the thermal coupling between the X-ray absorber and the bilayer, with a particular focus on the diameter of the two stems that connect the absorber to the TES bilayer, and the spacing between the stems and the bilayer.

Based on measurements of the noise, we found that the noise is reduced for shorter arms, an observation that we explain in terms of a higher internal thermal conductance between the absorber and the TES. Additionally, we observe that the position of kinks in the superconducting transition is affected by the spacing between the stems. While a more thorough investigation is required to fully understand the responsible mechanism, the effect appears to be related to previously reported observations in devices with added normal-metal structures on top of the bilayer. Comparing the devices with different stem diameters resulted in the striking observation that the smoothness of the transition seems to be significantly improved for the smaller stem diameter. Testing of additional devices of different geometry should help to confirm this observation. These devices are currently under fabrication.

The conclusion of this study is that the stems should be placed as close to the bilayer as possible in order to reduce the internal thermal fluctuation noise, leading to an improved energy resolution of up to 0.4$~$eV. This observation is consistent with the results from similar studies \cite{Wakeham2020}. The optimization of the stem diameter is more complicated. For the energy resolution, increasing the diameter could further reduce the ITFN and give better energy resolutions in certain bias points (even though the effect will be quite modest). However, our results suggest that reducing the stem diameter, or perhaps only using a single supporting stem, might greatly improve the smoothness and uniformity of the normal-to-superconducting transition, a feature that is of great importance for the successful operation of large number of TESs in any multiplexing scheme.

\section{Acknowledgement}
SRON is supported financially by NWO, the Netherlands Organization for Scientific Research. This work is partly funded by European Space Agency (ESA) and coordinated with other European efforts under ESA CTP contract ITT AO/1-7947/14/NL/BW.

\section{Data Availability}
The data that support the findings of this study are available from the corresponding author upon reasonable request.

\bibliography{ACG_bibliography}

\begin{thebibliography}{45}%
\makeatletter
\providecommand \@ifxundefined [1]{%
 \@ifx{#1\undefined}
}%
\providecommand \@ifnum [1]{%
 \ifnum #1\expandafter \@firstoftwo
 \else \expandafter \@secondoftwo
 \fi
}%
\providecommand \@ifx [1]{%
 \ifx #1\expandafter \@firstoftwo
 \else \expandafter \@secondoftwo
 \fi
}%
\providecommand \natexlab [1]{#1}%
\providecommand \enquote  [1]{``#1''}%
\providecommand \bibnamefont  [1]{#1}%
\providecommand \bibfnamefont [1]{#1}%
\providecommand \citenamefont [1]{#1}%
\providecommand \href@noop [0]{\@secondoftwo}%
\providecommand \href [0]{\begingroup \@sanitize@url \@href}%
\providecommand \@href[1]{\@@startlink{#1}\@@href}%
\providecommand \@@href[1]{\endgroup#1\@@endlink}%
\providecommand \@sanitize@url [0]{\catcode `\\12\catcode `\$12\catcode
  `\&12\catcode `\#12\catcode `\^12\catcode `\_12\catcode `\%12\relax}%
\providecommand \@@startlink[1]{}%
\providecommand \@@endlink[0]{}%
\providecommand \url  [0]{\begingroup\@sanitize@url \@url }%
\providecommand \@url [1]{\endgroup\@href {#1}{\urlprefix }}%
\providecommand \urlprefix  [0]{URL }%
\providecommand \Eprint [0]{\href }%
\providecommand \doibase [0]{https://doi.org/}%
\providecommand \selectlanguage [0]{\@gobble}%
\providecommand \bibinfo  [0]{\@secondoftwo}%
\providecommand \bibfield  [0]{\@secondoftwo}%
\providecommand \translation [1]{[#1]}%
\providecommand \BibitemOpen [0]{}%
\providecommand \bibitemStop [0]{}%
\providecommand \bibitemNoStop [0]{.\EOS\space}%
\providecommand \EOS [0]{\spacefactor3000\relax}%
\providecommand \BibitemShut  [1]{\csname bibitem#1\endcsname}%
\let\auto@bib@innerbib\@empty
\bibitem [{\citenamefont {Irwin}\ and\ \citenamefont
  {Hilton}(2005)}]{Irwin2005}%
  \BibitemOpen
  \bibfield  {author} {\bibinfo {author} {\bibfnamefont {K.}~\bibnamefont
  {Irwin}}\ and\ \bibinfo {author} {\bibfnamefont {G.}~\bibnamefont {Hilton}},\
  }\bibfield  {title} {\bibinfo {title} {{Transition-Edge Sensors}},\ }in\
  \href {https://doi.org/10.1007/10933596_3} {\emph {\bibinfo {booktitle}
  {Topics Appl. Phys.}}},\ Vol.~\bibinfo {volume} {99}\ (\bibinfo {year}
  {2005})\ pp.\ \bibinfo {pages} {63--150}\BibitemShut {NoStop}%
\bibitem [{\citenamefont {Ullom}\ and\ \citenamefont
  {Bennett}(2015)}]{Ullom2015}%
  \BibitemOpen
  \bibfield  {author} {\bibinfo {author} {\bibfnamefont {J.~N.}\ \bibnamefont
  {Ullom}}\ and\ \bibinfo {author} {\bibfnamefont {D.~A.}\ \bibnamefont
  {Bennett}},\ }\bibfield  {title} {\bibinfo {title} {{Review of
  superconducting transition-edge sensors for x-ray and gamma-ray
  spectroscopy}},\ }\href {https://doi.org/10.1088/0953-2048/28/8/084003}
  {\bibfield  {journal} {\bibinfo  {journal} {Superconductor Science and
  Technology}\ }\textbf {\bibinfo {volume} {28}},\ \bibinfo {pages} {084003}
  (\bibinfo {year} {2015})}\BibitemShut {NoStop}%
\bibitem [{\citenamefont {Gottardi}\ and\ \citenamefont
  {Nagayashi}(2021)}]{Gottardi2021a}%
  \BibitemOpen
  \bibfield  {author} {\bibinfo {author} {\bibfnamefont {L.}~\bibnamefont
  {Gottardi}}\ and\ \bibinfo {author} {\bibfnamefont {K.}~\bibnamefont
  {Nagayashi}},\ }\bibfield  {title} {\bibinfo {title} {{A review of X-ray
  microcalorimeters based on superconducting transition edge sensors for
  astrophysics and particle physics}},\ }\bibfield  {journal} {\bibinfo
  {journal} {Applied Sciences (Switzerland)}\ }\textbf {\bibinfo {volume}
  {11}},\ \href {https://doi.org/10.3390/app11093793} {10.3390/app11093793}
  (\bibinfo {year} {2021})\BibitemShut {NoStop}%
\bibitem [{\citenamefont {Barret}\ \emph {et~al.}(2020)\citenamefont {Barret},
  \citenamefont {Decourchelle}, \citenamefont {Fabian}, \citenamefont
  {Guainazzi}, \citenamefont {Nandra}, \citenamefont {Smith},\ and\
  \citenamefont {den Herder}}]{Barret2020a}%
  \BibitemOpen
  \bibfield  {author} {\bibinfo {author} {\bibfnamefont {D.}~\bibnamefont
  {Barret}}, \bibinfo {author} {\bibfnamefont {A.}~\bibnamefont
  {Decourchelle}}, \bibinfo {author} {\bibfnamefont {A.}~\bibnamefont
  {Fabian}}, \bibinfo {author} {\bibfnamefont {M.}~\bibnamefont {Guainazzi}},
  \bibinfo {author} {\bibfnamefont {K.}~\bibnamefont {Nandra}}, \bibinfo
  {author} {\bibfnamefont {R.}~\bibnamefont {Smith}},\ and\ \bibinfo {author}
  {\bibfnamefont {J.-W.}\ \bibnamefont {den Herder}},\ }\bibfield  {title}
  {\bibinfo {title} {{The Athena space X-ray observatory and the astrophysics
  of hot plasma†}},\ }\href {https://doi.org/10.1002/asna.202023782}
  {\bibfield  {journal} {\bibinfo  {journal} {Astronomische Nachrichten}\
  }\textbf {\bibinfo {volume} {341}},\ \bibinfo {pages} {224} (\bibinfo {year}
  {2020})},\ \Eprint {https://arxiv.org/abs/1912.04615} {arXiv:1912.04615}
  \BibitemShut {NoStop}%
\bibitem [{\citenamefont {Gaskin}\ and\ \citenamefont
  {Swartz}(2019)}]{Gaskin2019}%
  \BibitemOpen
  \bibfield  {author} {\bibinfo {author} {\bibfnamefont {J.~A.}\ \bibnamefont
  {Gaskin}}\ and\ \bibinfo {author} {\bibfnamefont {D.~A.}\ \bibnamefont
  {Swartz}},\ }\bibfield  {title} {\bibinfo {title} {{Lynx X-Ray Observatory:
  an overview}},\ }\href {https://doi.org/10.1117/1.JATIS.5.2.021001}
  {\bibfield  {journal} {\bibinfo  {journal} {Journal of Astronomical
  Telescopes, Instruments, and Systems}\ }\textbf {\bibinfo {volume} {5}},\
  \bibinfo {pages} {1} (\bibinfo {year} {2019})}\BibitemShut {NoStop}%
\bibitem [{\citenamefont {Cui}\ \emph {et~al.}(2020)\citenamefont {Cui},
  \citenamefont {Chen}, \citenamefont {Gao}, \citenamefont {Guo}, \citenamefont
  {Jin}, \citenamefont {Wang}, \citenamefont {Wang}, \citenamefont {Wang},
  \citenamefont {Wang}, \citenamefont {Wang}, \citenamefont {Wang},
  \citenamefont {Yuan},\ and\ \citenamefont {Zhang}}]{Cui2020}%
  \BibitemOpen
  \bibfield  {author} {\bibinfo {author} {\bibfnamefont {W.}~\bibnamefont
  {Cui}}, \bibinfo {author} {\bibfnamefont {L.~B.}\ \bibnamefont {Chen}},
  \bibinfo {author} {\bibfnamefont {B.}~\bibnamefont {Gao}}, \bibinfo {author}
  {\bibfnamefont {F.~L.}\ \bibnamefont {Guo}}, \bibinfo {author} {\bibfnamefont
  {H.}~\bibnamefont {Jin}}, \bibinfo {author} {\bibfnamefont {G.~L.}\
  \bibnamefont {Wang}}, \bibinfo {author} {\bibfnamefont {L.}~\bibnamefont
  {Wang}}, \bibinfo {author} {\bibfnamefont {J.~J.}\ \bibnamefont {Wang}},
  \bibinfo {author} {\bibfnamefont {W.}~\bibnamefont {Wang}}, \bibinfo {author}
  {\bibfnamefont {Z.~S.}\ \bibnamefont {Wang}}, \bibinfo {author}
  {\bibfnamefont {Z.}~\bibnamefont {Wang}}, \bibinfo {author} {\bibfnamefont
  {F.}~\bibnamefont {Yuan}},\ and\ \bibinfo {author} {\bibfnamefont
  {W.}~\bibnamefont {Zhang}},\ }\bibfield  {title} {\bibinfo {title} {{HUBS:
  Hot Universe Baryon Surveyor}},\ }\href
  {https://doi.org/10.1007/s10909-019-02279-3} {\bibfield  {journal} {\bibinfo
  {journal} {Journal of Low Temperature Physics}\ }\textbf {\bibinfo {volume}
  {199}},\ \bibinfo {pages} {502} (\bibinfo {year} {2020})}\BibitemShut
  {NoStop}%
\bibitem [{\citenamefont {Durkin}\ \emph {et~al.}(2019)\citenamefont {Durkin},
  \citenamefont {Adams}, \citenamefont {Bandler}, \citenamefont {Chervenak},
  \citenamefont {Chaudhuri}, \citenamefont {Dawson}, \citenamefont {Denison},
  \citenamefont {Doriese}, \citenamefont {Duff}, \citenamefont {Finkbeiner},
  \citenamefont {Fitzgerald}, \citenamefont {Fowler}, \citenamefont {Gard},
  \citenamefont {Hilton}, \citenamefont {Irwin}, \citenamefont {Joe},
  \citenamefont {Kelley}, \citenamefont {Kilbourne}, \citenamefont {Miniussi},
  \citenamefont {Morgan}, \citenamefont {O'Neil}, \citenamefont {Pappas},
  \citenamefont {Porter}, \citenamefont {Reintsema}, \citenamefont {Rudman},
  \citenamefont {Sakai}, \citenamefont {Smith}, \citenamefont {Stevens},
  \citenamefont {Swetz}, \citenamefont {Szypryt}, \citenamefont {Ullom},
  \citenamefont {Vale}, \citenamefont {Wakeham}, \citenamefont {Weber},\ and\
  \citenamefont {Young}}]{Durkin2019}%
  \BibitemOpen
  \bibfield  {author} {\bibinfo {author} {\bibfnamefont {M.}~\bibnamefont
  {Durkin}}, \bibinfo {author} {\bibfnamefont {J.~S.}\ \bibnamefont {Adams}},
  \bibinfo {author} {\bibfnamefont {S.~R.}\ \bibnamefont {Bandler}}, \bibinfo
  {author} {\bibfnamefont {J.~A.}\ \bibnamefont {Chervenak}}, \bibinfo {author}
  {\bibfnamefont {S.}~\bibnamefont {Chaudhuri}}, \bibinfo {author}
  {\bibfnamefont {C.~S.}\ \bibnamefont {Dawson}}, \bibinfo {author}
  {\bibfnamefont {E.~V.}\ \bibnamefont {Denison}}, \bibinfo {author}
  {\bibfnamefont {W.~B.}\ \bibnamefont {Doriese}}, \bibinfo {author}
  {\bibfnamefont {S.~M.}\ \bibnamefont {Duff}}, \bibinfo {author}
  {\bibfnamefont {F.~M.}\ \bibnamefont {Finkbeiner}}, \bibinfo {author}
  {\bibfnamefont {C.~T.}\ \bibnamefont {Fitzgerald}}, \bibinfo {author}
  {\bibfnamefont {J.~W.}\ \bibnamefont {Fowler}}, \bibinfo {author}
  {\bibfnamefont {J.~D.}\ \bibnamefont {Gard}}, \bibinfo {author}
  {\bibfnamefont {G.~C.}\ \bibnamefont {Hilton}}, \bibinfo {author}
  {\bibfnamefont {K.~D.}\ \bibnamefont {Irwin}}, \bibinfo {author}
  {\bibfnamefont {Y.~I.}\ \bibnamefont {Joe}}, \bibinfo {author} {\bibfnamefont
  {R.~L.}\ \bibnamefont {Kelley}}, \bibinfo {author} {\bibfnamefont {C.~A.}\
  \bibnamefont {Kilbourne}}, \bibinfo {author} {\bibfnamefont {A.~R.}\
  \bibnamefont {Miniussi}}, \bibinfo {author} {\bibfnamefont {K.~M.}\
  \bibnamefont {Morgan}}, \bibinfo {author} {\bibfnamefont {G.~C.}\
  \bibnamefont {O'Neil}}, \bibinfo {author} {\bibfnamefont {C.~G.}\
  \bibnamefont {Pappas}}, \bibinfo {author} {\bibfnamefont {F.~S.}\
  \bibnamefont {Porter}}, \bibinfo {author} {\bibfnamefont {C.~D.}\
  \bibnamefont {Reintsema}}, \bibinfo {author} {\bibfnamefont {D.~A.}\
  \bibnamefont {Rudman}}, \bibinfo {author} {\bibfnamefont {K.}~\bibnamefont
  {Sakai}}, \bibinfo {author} {\bibfnamefont {S.~J.}\ \bibnamefont {Smith}},
  \bibinfo {author} {\bibfnamefont {R.~W.}\ \bibnamefont {Stevens}}, \bibinfo
  {author} {\bibfnamefont {D.~S.}\ \bibnamefont {Swetz}}, \bibinfo {author}
  {\bibfnamefont {P.}~\bibnamefont {Szypryt}}, \bibinfo {author} {\bibfnamefont
  {J.~N.}\ \bibnamefont {Ullom}}, \bibinfo {author} {\bibfnamefont {L.~R.}\
  \bibnamefont {Vale}}, \bibinfo {author} {\bibfnamefont {N.~A.}\ \bibnamefont
  {Wakeham}}, \bibinfo {author} {\bibfnamefont {J.~C.}\ \bibnamefont {Weber}},\
  and\ \bibinfo {author} {\bibfnamefont {B.~A.}\ \bibnamefont {Young}},\
  }\bibfield  {title} {\bibinfo {title} {{Demonstration of athena X-IFU
  compatible 40-row time-division-multiplexed readout}},\ }\href
  {https://doi.org/10.1109/TASC.2019.2904472} {\bibfield  {journal} {\bibinfo
  {journal} {IEEE Transactions on Applied Superconductivity}\ }\textbf
  {\bibinfo {volume} {29}},\ \bibinfo {pages} {1} (\bibinfo {year}
  {2019})}\BibitemShut {NoStop}%
\bibitem [{\citenamefont {Akamatsu}\ \emph {et~al.}(2020)\citenamefont
  {Akamatsu}, \citenamefont {Gottardi}, \citenamefont {van~der Kuur},
  \citenamefont {de~Vries}, \citenamefont {Bruijn}, \citenamefont {Chervenak},
  \citenamefont {Kiviranta}, \citenamefont {van~den Linden}, \citenamefont
  {Jackson}, \citenamefont {Miniussi}, \citenamefont {Ravensberg},
  \citenamefont {Sakai}, \citenamefont {Smith},\ and\ \citenamefont
  {Wakeham}}]{Akamatsu2020}%
  \BibitemOpen
  \bibfield  {author} {\bibinfo {author} {\bibfnamefont {H.}~\bibnamefont
  {Akamatsu}}, \bibinfo {author} {\bibfnamefont {L.}~\bibnamefont {Gottardi}},
  \bibinfo {author} {\bibfnamefont {J.}~\bibnamefont {van~der Kuur}}, \bibinfo
  {author} {\bibfnamefont {C.~P.}\ \bibnamefont {de~Vries}}, \bibinfo {author}
  {\bibfnamefont {M.~P.}\ \bibnamefont {Bruijn}}, \bibinfo {author}
  {\bibfnamefont {J.~A.}\ \bibnamefont {Chervenak}}, \bibinfo {author}
  {\bibfnamefont {M.}~\bibnamefont {Kiviranta}}, \bibinfo {author}
  {\bibfnamefont {A.~J.}\ \bibnamefont {van~den Linden}}, \bibinfo {author}
  {\bibfnamefont {B.~D.}\ \bibnamefont {Jackson}}, \bibinfo {author}
  {\bibfnamefont {A.~R.}\ \bibnamefont {Miniussi}}, \bibinfo {author}
  {\bibfnamefont {K.}~\bibnamefont {Ravensberg}}, \bibinfo {author}
  {\bibfnamefont {K.}~\bibnamefont {Sakai}}, \bibinfo {author} {\bibfnamefont
  {S.~J.}\ \bibnamefont {Smith}},\ and\ \bibinfo {author} {\bibfnamefont
  {N.}~\bibnamefont {Wakeham}},\ }\bibfield  {title} {\bibinfo {title}
  {{Progress in the Development of Frequency-Domain Multiplexing for the X-ray
  Integral Field Unit on Board the Athena Mission}},\ }\href
  {https://doi.org/10.1007/s10909-020-02351-3} {\bibfield  {journal} {\bibinfo
  {journal} {Journal of Low Temperature Physics}\ }\textbf {\bibinfo {volume}
  {199}},\ \bibinfo {pages} {737} (\bibinfo {year} {2020})}\BibitemShut
  {NoStop}%
\bibitem [{\citenamefont {Morgan}\ \emph {et~al.}(2016)\citenamefont {Morgan},
  \citenamefont {Alpert}, \citenamefont {Bennett}, \citenamefont {Denison},
  \citenamefont {Doriese}, \citenamefont {Fowler}, \citenamefont {Gard},
  \citenamefont {Hilton}, \citenamefont {Irwin}, \citenamefont {Joe},
  \citenamefont {O'Neil}, \citenamefont {Reintsema}, \citenamefont {Schmidt},
  \citenamefont {Ullom},\ and\ \citenamefont {Swetz}}]{Morgan2016}%
  \BibitemOpen
  \bibfield  {author} {\bibinfo {author} {\bibfnamefont {K.~M.}\ \bibnamefont
  {Morgan}}, \bibinfo {author} {\bibfnamefont {B.~K.}\ \bibnamefont {Alpert}},
  \bibinfo {author} {\bibfnamefont {D.~A.}\ \bibnamefont {Bennett}}, \bibinfo
  {author} {\bibfnamefont {E.~V.}\ \bibnamefont {Denison}}, \bibinfo {author}
  {\bibfnamefont {W.~B.}\ \bibnamefont {Doriese}}, \bibinfo {author}
  {\bibfnamefont {J.~W.}\ \bibnamefont {Fowler}}, \bibinfo {author}
  {\bibfnamefont {J.~D.}\ \bibnamefont {Gard}}, \bibinfo {author}
  {\bibfnamefont {G.~C.}\ \bibnamefont {Hilton}}, \bibinfo {author}
  {\bibfnamefont {K.~D.}\ \bibnamefont {Irwin}}, \bibinfo {author}
  {\bibfnamefont {Y.~I.}\ \bibnamefont {Joe}}, \bibinfo {author} {\bibfnamefont
  {G.~C.}\ \bibnamefont {O'Neil}}, \bibinfo {author} {\bibfnamefont {C.~D.}\
  \bibnamefont {Reintsema}}, \bibinfo {author} {\bibfnamefont {D.~R.}\
  \bibnamefont {Schmidt}}, \bibinfo {author} {\bibfnamefont {J.~N.}\
  \bibnamefont {Ullom}},\ and\ \bibinfo {author} {\bibfnamefont {D.~S.}\
  \bibnamefont {Swetz}},\ }\bibfield  {title} {\bibinfo {title}
  {{Code-division-multiplexed readout of large arrays of TES
  microcalorimeters}},\ }\href {https://doi.org/10.1063/1.4962636} {\bibfield
  {journal} {\bibinfo  {journal} {Applied Physics Letters}\ }\textbf {\bibinfo
  {volume} {109}},\ \bibinfo {pages} {112604} (\bibinfo {year}
  {2016})}\BibitemShut {NoStop}%
\bibitem [{\citenamefont {Bennett}\ \emph {et~al.}(2019)\citenamefont
  {Bennett}, \citenamefont {Mates}, \citenamefont {Bandler}, \citenamefont
  {Becker}, \citenamefont {Fowler}, \citenamefont {Gard}, \citenamefont
  {Hilton}, \citenamefont {Irwin}, \citenamefont {Morgan}, \citenamefont
  {Reintsema}, \citenamefont {Sakai}, \citenamefont {Schmidt}, \citenamefont
  {Smith}, \citenamefont {Swetz}, \citenamefont {Ullom}, \citenamefont {Vale},\
  and\ \citenamefont {Wessels}}]{Bennett2019}%
  \BibitemOpen
  \bibfield  {author} {\bibinfo {author} {\bibfnamefont {D.~A.}\ \bibnamefont
  {Bennett}}, \bibinfo {author} {\bibfnamefont {J.~A.~B.}\ \bibnamefont
  {Mates}}, \bibinfo {author} {\bibfnamefont {S.~R.}\ \bibnamefont {Bandler}},
  \bibinfo {author} {\bibfnamefont {D.~T.}\ \bibnamefont {Becker}}, \bibinfo
  {author} {\bibfnamefont {J.~W.}\ \bibnamefont {Fowler}}, \bibinfo {author}
  {\bibfnamefont {J.~D.}\ \bibnamefont {Gard}}, \bibinfo {author}
  {\bibfnamefont {G.~C.}\ \bibnamefont {Hilton}}, \bibinfo {author}
  {\bibfnamefont {K.~D.}\ \bibnamefont {Irwin}}, \bibinfo {author}
  {\bibfnamefont {K.~M.}\ \bibnamefont {Morgan}}, \bibinfo {author}
  {\bibfnamefont {C.~D.}\ \bibnamefont {Reintsema}}, \bibinfo {author}
  {\bibfnamefont {K.}~\bibnamefont {Sakai}}, \bibinfo {author} {\bibfnamefont
  {D.~R.}\ \bibnamefont {Schmidt}}, \bibinfo {author} {\bibfnamefont {S.~J.}\
  \bibnamefont {Smith}}, \bibinfo {author} {\bibfnamefont {D.~S.}\ \bibnamefont
  {Swetz}}, \bibinfo {author} {\bibfnamefont {J.~N.}\ \bibnamefont {Ullom}},
  \bibinfo {author} {\bibfnamefont {L.~R.}\ \bibnamefont {Vale}},\ and\
  \bibinfo {author} {\bibfnamefont {A.~L.}\ \bibnamefont {Wessels}},\
  }\bibfield  {title} {\bibinfo {title} {{Microwave SQUID multiplexing for the
  Lynx x-ray microcalorimeter}},\ }\href
  {https://doi.org/10.1117/1.JATIS.5.2.021007} {\bibfield  {journal} {\bibinfo
  {journal} {Journal of Astronomical Telescopes, Instruments, and Systems}\
  }\textbf {\bibinfo {volume} {5}},\ \bibinfo {pages} {1} (\bibinfo {year}
  {2019})}\BibitemShut {NoStop}%
\bibitem [{\citenamefont {Nakashima}\ \emph {et~al.}(2020)\citenamefont
  {Nakashima}, \citenamefont {Hirayama}, \citenamefont {Kohjiro}, \citenamefont
  {Yamamori}, \citenamefont {Nagasawa}, \citenamefont {Sato}, \citenamefont
  {Yamada}, \citenamefont {Hayakawa}, \citenamefont {Yamasaki}, \citenamefont
  {Mitsuda}, \citenamefont {Nagayoshi}, \citenamefont {Akamatsu}, \citenamefont
  {Gottardi}, \citenamefont {Taralli}, \citenamefont {Bruijn}, \citenamefont
  {Ridder}, \citenamefont {Gao},\ and\ \citenamefont {{Den
  Herder}}}]{Nakashima2020}%
  \BibitemOpen
  \bibfield  {author} {\bibinfo {author} {\bibfnamefont {Y.}~\bibnamefont
  {Nakashima}}, \bibinfo {author} {\bibfnamefont {F.}~\bibnamefont {Hirayama}},
  \bibinfo {author} {\bibfnamefont {S.}~\bibnamefont {Kohjiro}}, \bibinfo
  {author} {\bibfnamefont {H.}~\bibnamefont {Yamamori}}, \bibinfo {author}
  {\bibfnamefont {S.}~\bibnamefont {Nagasawa}}, \bibinfo {author}
  {\bibfnamefont {A.}~\bibnamefont {Sato}}, \bibinfo {author} {\bibfnamefont
  {S.}~\bibnamefont {Yamada}}, \bibinfo {author} {\bibfnamefont
  {R.}~\bibnamefont {Hayakawa}}, \bibinfo {author} {\bibfnamefont {N.~Y.}\
  \bibnamefont {Yamasaki}}, \bibinfo {author} {\bibfnamefont {K.}~\bibnamefont
  {Mitsuda}}, \bibinfo {author} {\bibfnamefont {K.}~\bibnamefont {Nagayoshi}},
  \bibinfo {author} {\bibfnamefont {H.}~\bibnamefont {Akamatsu}}, \bibinfo
  {author} {\bibfnamefont {L.}~\bibnamefont {Gottardi}}, \bibinfo {author}
  {\bibfnamefont {E.}~\bibnamefont {Taralli}}, \bibinfo {author} {\bibfnamefont
  {M.~P.}\ \bibnamefont {Bruijn}}, \bibinfo {author} {\bibfnamefont {M.~L.}\
  \bibnamefont {Ridder}}, \bibinfo {author} {\bibfnamefont {J.~R.}\
  \bibnamefont {Gao}},\ and\ \bibinfo {author} {\bibfnamefont {J.~W.}\
  \bibnamefont {{Den Herder}}},\ }\bibfield  {title} {\bibinfo {title}
  {{Low-noise microwave SQUID multiplexed readout of 38 x-ray transition-edge
  sensor microcalorimeters}},\ }\href {https://doi.org/10.1063/5.0016333}
  {\bibfield  {journal} {\bibinfo  {journal} {Applied Physics Letters}\
  }\textbf {\bibinfo {volume} {117}},\ \bibinfo {pages} {3} (\bibinfo {year}
  {2020})}\BibitemShut {NoStop}%
\bibitem [{\citenamefont {Pajot}\ \emph {et~al.}(2018)\citenamefont {Pajot},
  \citenamefont {Barret}, \citenamefont {Lam-Trong}, \citenamefont {den
  Herder}, \citenamefont {Piro}, \citenamefont {Cappi}, \citenamefont
  {Huovelin}, \citenamefont {Kelley}, \citenamefont {Mas-Hesse}, \citenamefont
  {Mitsuda}, \citenamefont {Paltani}, \citenamefont {Rauw}, \citenamefont
  {Rozanska}, \citenamefont {Wilms}, \citenamefont {Barbera}, \citenamefont
  {Douchin}, \citenamefont {Geoffray}, \citenamefont {den Hartog},
  \citenamefont {Kilbourne}, \citenamefont {{Le Du}}, \citenamefont {Macculi},
  \citenamefont {Mesnager},\ and\ \citenamefont {Peille}}]{Pajot2018}%
  \BibitemOpen
  \bibfield  {author} {\bibinfo {author} {\bibfnamefont {F.}~\bibnamefont
  {Pajot}}, \bibinfo {author} {\bibfnamefont {D.}~\bibnamefont {Barret}},
  \bibinfo {author} {\bibfnamefont {T.}~\bibnamefont {Lam-Trong}}, \bibinfo
  {author} {\bibfnamefont {J.-W.}\ \bibnamefont {den Herder}}, \bibinfo
  {author} {\bibfnamefont {L.}~\bibnamefont {Piro}}, \bibinfo {author}
  {\bibfnamefont {M.}~\bibnamefont {Cappi}}, \bibinfo {author} {\bibfnamefont
  {J.}~\bibnamefont {Huovelin}}, \bibinfo {author} {\bibfnamefont
  {R.}~\bibnamefont {Kelley}}, \bibinfo {author} {\bibfnamefont {J.~M.}\
  \bibnamefont {Mas-Hesse}}, \bibinfo {author} {\bibfnamefont {K.}~\bibnamefont
  {Mitsuda}}, \bibinfo {author} {\bibfnamefont {S.}~\bibnamefont {Paltani}},
  \bibinfo {author} {\bibfnamefont {G.}~\bibnamefont {Rauw}}, \bibinfo {author}
  {\bibfnamefont {A.}~\bibnamefont {Rozanska}}, \bibinfo {author}
  {\bibfnamefont {J.}~\bibnamefont {Wilms}}, \bibinfo {author} {\bibfnamefont
  {M.}~\bibnamefont {Barbera}}, \bibinfo {author} {\bibfnamefont
  {F.}~\bibnamefont {Douchin}}, \bibinfo {author} {\bibfnamefont
  {H.}~\bibnamefont {Geoffray}}, \bibinfo {author} {\bibfnamefont {R.~H.}\
  \bibnamefont {den Hartog}}, \bibinfo {author} {\bibfnamefont
  {C.}~\bibnamefont {Kilbourne}}, \bibinfo {author} {\bibfnamefont
  {M.}~\bibnamefont {{Le Du}}}, \bibinfo {author} {\bibfnamefont
  {C.}~\bibnamefont {Macculi}}, \bibinfo {author} {\bibfnamefont {J.~M.}\
  \bibnamefont {Mesnager}},\ and\ \bibinfo {author} {\bibfnamefont
  {P.}~\bibnamefont {Peille}},\ }\bibfield  {title} {\bibinfo {title} {{The
  Athena X-ray Integral Field Unit (X-IFU)}},\ }\href
  {https://doi.org/10.1007/s10909-018-1904-5} {\bibfield  {journal} {\bibinfo
  {journal} {Journal of Low Temperature Physics}\ }\textbf {\bibinfo {volume}
  {193}},\ \bibinfo {pages} {901} (\bibinfo {year} {2018})}\BibitemShut
  {NoStop}%
\bibitem [{\citenamefont {Hays-Wehle}\ \emph {et~al.}(2016)\citenamefont
  {Hays-Wehle}, \citenamefont {Schmidt}, \citenamefont {Ullom},\ and\
  \citenamefont {Swetz}}]{Hays-Wehle2016}%
  \BibitemOpen
  \bibfield  {author} {\bibinfo {author} {\bibfnamefont {J.~P.}\ \bibnamefont
  {Hays-Wehle}}, \bibinfo {author} {\bibfnamefont {D.~R.}\ \bibnamefont
  {Schmidt}}, \bibinfo {author} {\bibfnamefont {J.~N.}\ \bibnamefont {Ullom}},\
  and\ \bibinfo {author} {\bibfnamefont {D.~S.}\ \bibnamefont {Swetz}},\
  }\bibfield  {title} {\bibinfo {title} {{Thermal Conductance Engineering for
  High-Speed TES Microcalorimeters}},\ }\href
  {https://doi.org/10.1007/s10909-015-1416-5} {\bibfield  {journal} {\bibinfo
  {journal} {Journal of Low Temperature Physics}\ }\textbf {\bibinfo {volume}
  {184}},\ \bibinfo {pages} {492} (\bibinfo {year} {2016})}\BibitemShut
  {NoStop}%
\bibitem [{\citenamefont {Morgan}\ \emph {et~al.}(2017)\citenamefont {Morgan},
  \citenamefont {Pappas}, \citenamefont {Bennett}, \citenamefont {Gard},
  \citenamefont {Hays-Wehle}, \citenamefont {Hilton}, \citenamefont
  {Reintsema}, \citenamefont {Schmidt}, \citenamefont {Ullom},\ and\
  \citenamefont {Swetz}}]{Morgan2017}%
  \BibitemOpen
  \bibfield  {author} {\bibinfo {author} {\bibfnamefont {K.~M.}\ \bibnamefont
  {Morgan}}, \bibinfo {author} {\bibfnamefont {C.~G.}\ \bibnamefont {Pappas}},
  \bibinfo {author} {\bibfnamefont {D.~A.}\ \bibnamefont {Bennett}}, \bibinfo
  {author} {\bibfnamefont {J.~D.}\ \bibnamefont {Gard}}, \bibinfo {author}
  {\bibfnamefont {J.~P.}\ \bibnamefont {Hays-Wehle}}, \bibinfo {author}
  {\bibfnamefont {G.~C.}\ \bibnamefont {Hilton}}, \bibinfo {author}
  {\bibfnamefont {C.~D.}\ \bibnamefont {Reintsema}}, \bibinfo {author}
  {\bibfnamefont {D.~R.}\ \bibnamefont {Schmidt}}, \bibinfo {author}
  {\bibfnamefont {J.~N.}\ \bibnamefont {Ullom}},\ and\ \bibinfo {author}
  {\bibfnamefont {D.~S.}\ \bibnamefont {Swetz}},\ }\bibfield  {title} {\bibinfo
  {title} {{Dependence of transition width on current and critical current in
  transition-edge sensors}},\ }\href {https://doi.org/10.1063/1.4984065}
  {\bibfield  {journal} {\bibinfo  {journal} {Applied Physics Letters}\
  }\textbf {\bibinfo {volume} {110}},\ \bibinfo {pages} {2} (\bibinfo {year}
  {2017})}\BibitemShut {NoStop}%
\bibitem [{\citenamefont {Zhang}\ \emph {et~al.}(2019)\citenamefont {Zhang},
  \citenamefont {Duff}, \citenamefont {Hilton}, \citenamefont {Lowell},
  \citenamefont {Morgan}, \citenamefont {Schmidt},\ and\ \citenamefont
  {Ullom}}]{Zhang2019}%
  \BibitemOpen
  \bibfield  {author} {\bibinfo {author} {\bibfnamefont {X.}~\bibnamefont
  {Zhang}}, \bibinfo {author} {\bibfnamefont {S.~M.}\ \bibnamefont {Duff}},
  \bibinfo {author} {\bibfnamefont {G.~C.}\ \bibnamefont {Hilton}}, \bibinfo
  {author} {\bibfnamefont {P.~J.}\ \bibnamefont {Lowell}}, \bibinfo {author}
  {\bibfnamefont {K.~M.}\ \bibnamefont {Morgan}}, \bibinfo {author}
  {\bibfnamefont {D.~R.}\ \bibnamefont {Schmidt}},\ and\ \bibinfo {author}
  {\bibfnamefont {J.~N.}\ \bibnamefont {Ullom}},\ }\bibfield  {title} {\bibinfo
  {title} {{Controlling the thermal conductance of silicon nitride membranes at
  100 mK temperatures with patterned metal features}},\ }\href
  {https://doi.org/10.1063/1.5097173} {\bibfield  {journal} {\bibinfo
  {journal} {Applied Physics Letters}\ }\textbf {\bibinfo {volume} {115}},\
  \bibinfo {pages} {052601} (\bibinfo {year} {2019})}\BibitemShut {NoStop}%
\bibitem [{\citenamefont {Gottardi}\ \emph {et~al.}(2017)\citenamefont
  {Gottardi}, \citenamefont {Akamatsu}, \citenamefont {van~der Kuur},
  \citenamefont {Smith}, \citenamefont {Kozorezov},\ and\ \citenamefont
  {Chervenak}}]{Gottardi2017}%
  \BibitemOpen
  \bibfield  {author} {\bibinfo {author} {\bibfnamefont {L.}~\bibnamefont
  {Gottardi}}, \bibinfo {author} {\bibfnamefont {H.}~\bibnamefont {Akamatsu}},
  \bibinfo {author} {\bibfnamefont {J.}~\bibnamefont {van~der Kuur}}, \bibinfo
  {author} {\bibfnamefont {S.~J.}\ \bibnamefont {Smith}}, \bibinfo {author}
  {\bibfnamefont {A.}~\bibnamefont {Kozorezov}},\ and\ \bibinfo {author}
  {\bibfnamefont {J.}~\bibnamefont {Chervenak}},\ }\bibfield  {title} {\bibinfo
  {title} {{Study of TES-Based Microcalorimeters of Different Size and Geometry
  Under AC Bias}},\ }\href {https://doi.org/10.1109/TASC.2017.2655500}
  {\bibfield  {journal} {\bibinfo  {journal} {IEEE Transactions on Applied
  Superconductivity}\ }\textbf {\bibinfo {volume} {27}},\ \bibinfo {pages} {1}
  (\bibinfo {year} {2017})}\BibitemShut {NoStop}%
\bibitem [{\citenamefont {Sakai}\ \emph {et~al.}(2018)\citenamefont {Sakai},
  \citenamefont {Adams}, \citenamefont {Bandler}, \citenamefont {Chervenak},
  \citenamefont {Datesman}, \citenamefont {Eckart}, \citenamefont {Finkbeiner},
  \citenamefont {Kelley}, \citenamefont {Kilbourne}, \citenamefont {Miniussi},
  \citenamefont {Porter}, \citenamefont {Sadleir}, \citenamefont {Smith},
  \citenamefont {Wakeham}, \citenamefont {Wassell}, \citenamefont {Yoon},
  \citenamefont {Akamatsu}, \citenamefont {Bruijn}, \citenamefont {Gottardi},
  \citenamefont {Jackson}, \citenamefont {van~der Kuur}, \citenamefont {van
  Leeuwen}, \citenamefont {van~der Linden}, \citenamefont {van Weers},\ and\
  \citenamefont {Kiviranta}}]{Sakai2018}%
  \BibitemOpen
  \bibfield  {author} {\bibinfo {author} {\bibfnamefont {K.}~\bibnamefont
  {Sakai}}, \bibinfo {author} {\bibfnamefont {J.~S.}\ \bibnamefont {Adams}},
  \bibinfo {author} {\bibfnamefont {S.~R.}\ \bibnamefont {Bandler}}, \bibinfo
  {author} {\bibfnamefont {J.~A.}\ \bibnamefont {Chervenak}}, \bibinfo {author}
  {\bibfnamefont {A.~M.}\ \bibnamefont {Datesman}}, \bibinfo {author}
  {\bibfnamefont {M.~E.}\ \bibnamefont {Eckart}}, \bibinfo {author}
  {\bibfnamefont {F.~M.}\ \bibnamefont {Finkbeiner}}, \bibinfo {author}
  {\bibfnamefont {R.~L.}\ \bibnamefont {Kelley}}, \bibinfo {author}
  {\bibfnamefont {C.~A.}\ \bibnamefont {Kilbourne}}, \bibinfo {author}
  {\bibfnamefont {A.~R.}\ \bibnamefont {Miniussi}}, \bibinfo {author}
  {\bibfnamefont {F.~S.}\ \bibnamefont {Porter}}, \bibinfo {author}
  {\bibfnamefont {J.~S.}\ \bibnamefont {Sadleir}}, \bibinfo {author}
  {\bibfnamefont {S.~J.}\ \bibnamefont {Smith}}, \bibinfo {author}
  {\bibfnamefont {N.~A.}\ \bibnamefont {Wakeham}}, \bibinfo {author}
  {\bibfnamefont {E.~J.}\ \bibnamefont {Wassell}}, \bibinfo {author}
  {\bibfnamefont {W.}~\bibnamefont {Yoon}}, \bibinfo {author} {\bibfnamefont
  {H.}~\bibnamefont {Akamatsu}}, \bibinfo {author} {\bibfnamefont {M.~P.}\
  \bibnamefont {Bruijn}}, \bibinfo {author} {\bibfnamefont {L.}~\bibnamefont
  {Gottardi}}, \bibinfo {author} {\bibfnamefont {B.~D.}\ \bibnamefont
  {Jackson}}, \bibinfo {author} {\bibfnamefont {J.}~\bibnamefont {van~der
  Kuur}}, \bibinfo {author} {\bibfnamefont {B.~J.}\ \bibnamefont {van
  Leeuwen}}, \bibinfo {author} {\bibfnamefont {A.~J.}\ \bibnamefont {van~der
  Linden}}, \bibinfo {author} {\bibfnamefont {H.~J.}\ \bibnamefont {van
  Weers}},\ and\ \bibinfo {author} {\bibfnamefont {M.}~\bibnamefont
  {Kiviranta}},\ }\bibfield  {title} {\bibinfo {title} {{Study of Dissipative
  Losses in AC-Biased Mo/Au Bilayer Transition-Edge Sensors}},\ }\href
  {https://doi.org/10.1007/s10909-018-2002-4} {\bibfield  {journal} {\bibinfo
  {journal} {Journal of Low Temperature Physics}\ }\textbf {\bibinfo {volume}
  {193}},\ \bibinfo {pages} {356} (\bibinfo {year} {2018})}\BibitemShut
  {NoStop}%
\bibitem [{\citenamefont {Morgan}\ \emph {et~al.}(2019)\citenamefont {Morgan},
  \citenamefont {Becker}, \citenamefont {Bennett}, \citenamefont {Doriese},
  \citenamefont {Gard}, \citenamefont {Irwin}, \citenamefont {Lee},
  \citenamefont {Li}, \citenamefont {Mates}, \citenamefont {Pappas},
  \citenamefont {Schmidt}, \citenamefont {Titus}, \citenamefont {{Van Winkle}},
  \citenamefont {Ullom}, \citenamefont {Wessels},\ and\ \citenamefont
  {Swetz}}]{Morgan2019}%
  \BibitemOpen
  \bibfield  {author} {\bibinfo {author} {\bibfnamefont {K.~M.}\ \bibnamefont
  {Morgan}}, \bibinfo {author} {\bibfnamefont {D.~T.}\ \bibnamefont {Becker}},
  \bibinfo {author} {\bibfnamefont {D.~A.}\ \bibnamefont {Bennett}}, \bibinfo
  {author} {\bibfnamefont {W.~B.}\ \bibnamefont {Doriese}}, \bibinfo {author}
  {\bibfnamefont {J.~D.}\ \bibnamefont {Gard}}, \bibinfo {author}
  {\bibfnamefont {K.~D.}\ \bibnamefont {Irwin}}, \bibinfo {author}
  {\bibfnamefont {S.~J.}\ \bibnamefont {Lee}}, \bibinfo {author} {\bibfnamefont
  {D.}~\bibnamefont {Li}}, \bibinfo {author} {\bibfnamefont {J.~A.}\
  \bibnamefont {Mates}}, \bibinfo {author} {\bibfnamefont {C.~G.}\ \bibnamefont
  {Pappas}}, \bibinfo {author} {\bibfnamefont {D.~R.}\ \bibnamefont {Schmidt}},
  \bibinfo {author} {\bibfnamefont {C.~J.}\ \bibnamefont {Titus}}, \bibinfo
  {author} {\bibfnamefont {D.~D.}\ \bibnamefont {{Van Winkle}}}, \bibinfo
  {author} {\bibfnamefont {J.~N.}\ \bibnamefont {Ullom}}, \bibinfo {author}
  {\bibfnamefont {A.}~\bibnamefont {Wessels}},\ and\ \bibinfo {author}
  {\bibfnamefont {D.~S.}\ \bibnamefont {Swetz}},\ }\bibfield  {title} {\bibinfo
  {title} {{Use of Transition Models to Design High Performance TESs for the
  LCLS-II Soft X-Ray Spectrometer}},\ }\href
  {https://doi.org/10.1109/TASC.2019.2903032} {\bibfield  {journal} {\bibinfo
  {journal} {IEEE Transactions on Applied Superconductivity}\ }\textbf
  {\bibinfo {volume} {29}},\ \bibinfo {pages} {1} (\bibinfo {year}
  {2019})}\BibitemShut {NoStop}%
\bibitem [{\citenamefont {{De Wit}}\ \emph {et~al.}(2020)\citenamefont {{De
  Wit}}, \citenamefont {Gottardi}, \citenamefont {Taralli}, \citenamefont
  {Nagayoshi}, \citenamefont {Ridder}, \citenamefont {Akamatsu}, \citenamefont
  {Bruijn}, \citenamefont {D'Andrea}, \citenamefont {{Van Der Kuur}},
  \citenamefont {Ravensberg}, \citenamefont {Vaccaro}, \citenamefont {Visser},
  \citenamefont {Gao},\ and\ \citenamefont {{Den Herder}}}]{DeWit2020}%
  \BibitemOpen
  \bibfield  {author} {\bibinfo {author} {\bibfnamefont {M.}~\bibnamefont {{De
  Wit}}}, \bibinfo {author} {\bibfnamefont {L.}~\bibnamefont {Gottardi}},
  \bibinfo {author} {\bibfnamefont {E.}~\bibnamefont {Taralli}}, \bibinfo
  {author} {\bibfnamefont {K.}~\bibnamefont {Nagayoshi}}, \bibinfo {author}
  {\bibfnamefont {M.~L.}\ \bibnamefont {Ridder}}, \bibinfo {author}
  {\bibfnamefont {H.}~\bibnamefont {Akamatsu}}, \bibinfo {author}
  {\bibfnamefont {M.~P.}\ \bibnamefont {Bruijn}}, \bibinfo {author}
  {\bibfnamefont {M.}~\bibnamefont {D'Andrea}}, \bibinfo {author}
  {\bibfnamefont {J.}~\bibnamefont {{Van Der Kuur}}}, \bibinfo {author}
  {\bibfnamefont {K.}~\bibnamefont {Ravensberg}}, \bibinfo {author}
  {\bibfnamefont {D.}~\bibnamefont {Vaccaro}}, \bibinfo {author} {\bibfnamefont
  {S.}~\bibnamefont {Visser}}, \bibinfo {author} {\bibfnamefont {J.~R.}\
  \bibnamefont {Gao}},\ and\ \bibinfo {author} {\bibfnamefont {J.~W.}\
  \bibnamefont {{Den Herder}}},\ }\bibfield  {title} {\bibinfo {title} {{High
  aspect ratio transition edge sensors for x-ray spectrometry}},\ }\bibfield
  {journal} {\bibinfo  {journal} {Journal of Applied Physics}\ }\textbf
  {\bibinfo {volume} {128}},\ \href {https://doi.org/10.1063/5.0029669}
  {10.1063/5.0029669} (\bibinfo {year} {2020})\BibitemShut {NoStop}%
\bibitem [{\citenamefont {Ullom}\ \emph {et~al.}(2004)\citenamefont {Ullom},
  \citenamefont {Doriese}, \citenamefont {Hilton}, \citenamefont {Beall},
  \citenamefont {Deiker}, \citenamefont {Duncan}, \citenamefont {Ferreira},
  \citenamefont {Irwin}, \citenamefont {Reintsema},\ and\ \citenamefont
  {Vale}}]{Ullom2004}%
  \BibitemOpen
  \bibfield  {author} {\bibinfo {author} {\bibfnamefont {J.~N.}\ \bibnamefont
  {Ullom}}, \bibinfo {author} {\bibfnamefont {W.~B.}\ \bibnamefont {Doriese}},
  \bibinfo {author} {\bibfnamefont {G.~C.}\ \bibnamefont {Hilton}}, \bibinfo
  {author} {\bibfnamefont {J.~A.}\ \bibnamefont {Beall}}, \bibinfo {author}
  {\bibfnamefont {S.}~\bibnamefont {Deiker}}, \bibinfo {author} {\bibfnamefont
  {W.~D.}\ \bibnamefont {Duncan}}, \bibinfo {author} {\bibfnamefont
  {L.}~\bibnamefont {Ferreira}}, \bibinfo {author} {\bibfnamefont {K.~D.}\
  \bibnamefont {Irwin}}, \bibinfo {author} {\bibfnamefont {C.~D.}\ \bibnamefont
  {Reintsema}},\ and\ \bibinfo {author} {\bibfnamefont {L.~R.}\ \bibnamefont
  {Vale}},\ }\bibfield  {title} {\bibinfo {title} {{Characterization and
  reduction of unexplained noise in superconducting transition-edge sensors}},\
  }\href {https://doi.org/10.1063/1.1753058} {\bibfield  {journal} {\bibinfo
  {journal} {Applied Physics Letters}\ }\textbf {\bibinfo {volume} {84}},\
  \bibinfo {pages} {4206} (\bibinfo {year} {2004})}\BibitemShut {NoStop}%
\bibitem [{\citenamefont {Sadleir}\ \emph {et~al.}(2011)\citenamefont
  {Sadleir}, \citenamefont {Smith}, \citenamefont {Robinson}, \citenamefont
  {Finkbeiner}, \citenamefont {Chervenak}, \citenamefont {Bandler},
  \citenamefont {Eckart},\ and\ \citenamefont {Kilbourne}}]{Sadleir2011}%
  \BibitemOpen
  \bibfield  {author} {\bibinfo {author} {\bibfnamefont {J.~E.}\ \bibnamefont
  {Sadleir}}, \bibinfo {author} {\bibfnamefont {S.~J.}\ \bibnamefont {Smith}},
  \bibinfo {author} {\bibfnamefont {I.~K.}\ \bibnamefont {Robinson}}, \bibinfo
  {author} {\bibfnamefont {F.~M.}\ \bibnamefont {Finkbeiner}}, \bibinfo
  {author} {\bibfnamefont {J.~A.}\ \bibnamefont {Chervenak}}, \bibinfo {author}
  {\bibfnamefont {S.~R.}\ \bibnamefont {Bandler}}, \bibinfo {author}
  {\bibfnamefont {M.~E.}\ \bibnamefont {Eckart}},\ and\ \bibinfo {author}
  {\bibfnamefont {C.~A.}\ \bibnamefont {Kilbourne}},\ }\bibfield  {title}
  {\bibinfo {title} {{Proximity effects and nonequilibrium superconductivity in
  transition-edge sensors}},\ }\href
  {https://doi.org/10.1103/PhysRevB.84.184502} {\bibfield  {journal} {\bibinfo
  {journal} {Physical Review B}\ }\textbf {\bibinfo {volume} {84}},\ \bibinfo
  {pages} {184502} (\bibinfo {year} {2011})}\BibitemShut {NoStop}%
\bibitem [{\citenamefont {Miniussi}\ \emph {et~al.}(2018)\citenamefont
  {Miniussi}, \citenamefont {Adams}, \citenamefont {Bandler}, \citenamefont
  {Chervenak}, \citenamefont {Datesman}, \citenamefont {Eckart}, \citenamefont
  {Ewin}, \citenamefont {Finkbeiner}, \citenamefont {Kelley}, \citenamefont
  {Kilbourne}, \citenamefont {Porter}, \citenamefont {Sadleir}, \citenamefont
  {Sakai}, \citenamefont {Smith}, \citenamefont {Wakeham}, \citenamefont
  {Wassell},\ and\ \citenamefont {Yoon}}]{Miniussi2018}%
  \BibitemOpen
  \bibfield  {author} {\bibinfo {author} {\bibfnamefont {A.~R.}\ \bibnamefont
  {Miniussi}}, \bibinfo {author} {\bibfnamefont {J.~S.}\ \bibnamefont {Adams}},
  \bibinfo {author} {\bibfnamefont {S.~R.}\ \bibnamefont {Bandler}}, \bibinfo
  {author} {\bibfnamefont {J.~A.}\ \bibnamefont {Chervenak}}, \bibinfo {author}
  {\bibfnamefont {A.~M.}\ \bibnamefont {Datesman}}, \bibinfo {author}
  {\bibfnamefont {M.~E.}\ \bibnamefont {Eckart}}, \bibinfo {author}
  {\bibfnamefont {A.~J.}\ \bibnamefont {Ewin}}, \bibinfo {author}
  {\bibfnamefont {F.~M.}\ \bibnamefont {Finkbeiner}}, \bibinfo {author}
  {\bibfnamefont {R.~L.}\ \bibnamefont {Kelley}}, \bibinfo {author}
  {\bibfnamefont {C.~A.}\ \bibnamefont {Kilbourne}}, \bibinfo {author}
  {\bibfnamefont {F.~S.}\ \bibnamefont {Porter}}, \bibinfo {author}
  {\bibfnamefont {J.~E.}\ \bibnamefont {Sadleir}}, \bibinfo {author}
  {\bibfnamefont {K.}~\bibnamefont {Sakai}}, \bibinfo {author} {\bibfnamefont
  {S.~J.}\ \bibnamefont {Smith}}, \bibinfo {author} {\bibfnamefont {N.~A.}\
  \bibnamefont {Wakeham}}, \bibinfo {author} {\bibfnamefont {E.~J.}\
  \bibnamefont {Wassell}},\ and\ \bibinfo {author} {\bibfnamefont
  {W.}~\bibnamefont {Yoon}},\ }\bibfield  {title} {\bibinfo {title}
  {{Performance of an X-ray Microcalorimeter with a 240 $\mu$m Absorber and a
  50 $\mu$m TES Bilayer}},\ }\href {https://doi.org/10.1007/s10909-018-1974-4}
  {\bibfield  {journal} {\bibinfo  {journal} {Journal of Low Temperature
  Physics}\ }\textbf {\bibinfo {volume} {193}},\ \bibinfo {pages} {337}
  (\bibinfo {year} {2018})}\BibitemShut {NoStop}%
\bibitem [{\citenamefont {Wakeham}\ \emph {et~al.}(2018)\citenamefont
  {Wakeham}, \citenamefont {Adams}, \citenamefont {Bandler}, \citenamefont
  {Chervenak}, \citenamefont {Datesman}, \citenamefont {Eckart}, \citenamefont
  {Finkbeiner}, \citenamefont {Kelley}, \citenamefont {Kilbourne},
  \citenamefont {Miniussi}, \citenamefont {Porter}, \citenamefont {Sadleir},
  \citenamefont {Sakai}, \citenamefont {Smith}, \citenamefont {Wassell},\ and\
  \citenamefont {Yoon}}]{Wakeham2018}%
  \BibitemOpen
  \bibfield  {author} {\bibinfo {author} {\bibfnamefont {N.~A.}\ \bibnamefont
  {Wakeham}}, \bibinfo {author} {\bibfnamefont {J.~S.}\ \bibnamefont {Adams}},
  \bibinfo {author} {\bibfnamefont {S.~R.}\ \bibnamefont {Bandler}}, \bibinfo
  {author} {\bibfnamefont {J.~A.}\ \bibnamefont {Chervenak}}, \bibinfo {author}
  {\bibfnamefont {A.~M.}\ \bibnamefont {Datesman}}, \bibinfo {author}
  {\bibfnamefont {M.~E.}\ \bibnamefont {Eckart}}, \bibinfo {author}
  {\bibfnamefont {F.~M.}\ \bibnamefont {Finkbeiner}}, \bibinfo {author}
  {\bibfnamefont {R.~L.}\ \bibnamefont {Kelley}}, \bibinfo {author}
  {\bibfnamefont {C.~A.}\ \bibnamefont {Kilbourne}}, \bibinfo {author}
  {\bibfnamefont {A.~R.}\ \bibnamefont {Miniussi}}, \bibinfo {author}
  {\bibfnamefont {F.~S.}\ \bibnamefont {Porter}}, \bibinfo {author}
  {\bibfnamefont {J.~E.}\ \bibnamefont {Sadleir}}, \bibinfo {author}
  {\bibfnamefont {K.}~\bibnamefont {Sakai}}, \bibinfo {author} {\bibfnamefont
  {S.~J.}\ \bibnamefont {Smith}}, \bibinfo {author} {\bibfnamefont {E.~J.}\
  \bibnamefont {Wassell}},\ and\ \bibinfo {author} {\bibfnamefont
  {W.}~\bibnamefont {Yoon}},\ }\bibfield  {title} {\bibinfo {title} {{Effects
  of Normal Metal Features on Superconducting Transition-Edge Sensors}},\
  }\href {https://doi.org/10.1007/s10909-018-1898-z} {\bibfield  {journal}
  {\bibinfo  {journal} {Journal of Low Temperature Physics}\ }\textbf {\bibinfo
  {volume} {193}},\ \bibinfo {pages} {231} (\bibinfo {year}
  {2018})}\BibitemShut {NoStop}%
\bibitem [{\citenamefont {Kilbourne}\ \emph {et~al.}(2007)\citenamefont
  {Kilbourne}, \citenamefont {Bandler}, \citenamefont {Brown}, \citenamefont
  {Chervenak}, \citenamefont {Figueroa-Feliciano}, \citenamefont {Finkbeiner},
  \citenamefont {Iyomoto}, \citenamefont {Kelley}, \citenamefont {Porter},\
  and\ \citenamefont {Smith}}]{Kilbourne2007a}%
  \BibitemOpen
  \bibfield  {author} {\bibinfo {author} {\bibfnamefont {C.~A.}\ \bibnamefont
  {Kilbourne}}, \bibinfo {author} {\bibfnamefont {S.~R.}\ \bibnamefont
  {Bandler}}, \bibinfo {author} {\bibfnamefont {A.~D.}\ \bibnamefont {Brown}},
  \bibinfo {author} {\bibfnamefont {J.~A.}\ \bibnamefont {Chervenak}}, \bibinfo
  {author} {\bibfnamefont {E.}~\bibnamefont {Figueroa-Feliciano}}, \bibinfo
  {author} {\bibfnamefont {F.~M.}\ \bibnamefont {Finkbeiner}}, \bibinfo
  {author} {\bibfnamefont {N.}~\bibnamefont {Iyomoto}}, \bibinfo {author}
  {\bibfnamefont {R.~L.}\ \bibnamefont {Kelley}}, \bibinfo {author}
  {\bibfnamefont {F.~S.}\ \bibnamefont {Porter}},\ and\ \bibinfo {author}
  {\bibfnamefont {S.~J.}\ \bibnamefont {Smith}},\ }\bibfield  {title} {\bibinfo
  {title} {{Uniform high spectral resolution demonstrated in arrays of TES
  x-ray microcalorimeters}},\ }\href {https://doi.org/10.1117/12.734830}
  {\bibfield  {journal} {\bibinfo  {journal} {UV, X-Ray, and Gamma-Ray Space
  Instrumentation for Astronomy XV}\ }\textbf {\bibinfo {volume} {6686}},\
  \bibinfo {pages} {668606} (\bibinfo {year} {2007})}\BibitemShut {NoStop}%
\bibitem [{\citenamefont {Smith}\ \emph {et~al.}(2012)\citenamefont {Smith},
  \citenamefont {Adams}, \citenamefont {Eckart}, \citenamefont {Bailey},
  \citenamefont {Bandler}, \citenamefont {Chervenak}, \citenamefont
  {Finkbeiner}, \citenamefont {Kelley}, \citenamefont {Kilbourne},
  \citenamefont {Porter},\ and\ \citenamefont {Sadleir}}]{Smith2012}%
  \BibitemOpen
  \bibfield  {author} {\bibinfo {author} {\bibfnamefont {S.~J.}\ \bibnamefont
  {Smith}}, \bibinfo {author} {\bibfnamefont {J.~S.}\ \bibnamefont {Adams}},
  \bibinfo {author} {\bibfnamefont {M.~E.}\ \bibnamefont {Eckart}}, \bibinfo
  {author} {\bibfnamefont {C.~N.}\ \bibnamefont {Bailey}}, \bibinfo {author}
  {\bibfnamefont {S.~R.}\ \bibnamefont {Bandler}}, \bibinfo {author}
  {\bibfnamefont {J.~A.}\ \bibnamefont {Chervenak}}, \bibinfo {author}
  {\bibfnamefont {F.~M.}\ \bibnamefont {Finkbeiner}}, \bibinfo {author}
  {\bibfnamefont {R.~L.}\ \bibnamefont {Kelley}}, \bibinfo {author}
  {\bibfnamefont {C.~A.}\ \bibnamefont {Kilbourne}}, \bibinfo {author}
  {\bibfnamefont {F.~S.}\ \bibnamefont {Porter}},\ and\ \bibinfo {author}
  {\bibfnamefont {J.~E.}\ \bibnamefont {Sadleir}},\ }\bibfield  {title}
  {\bibinfo {title} {{Small pitch transition-edge sensors with broadband high
  spectral resolution for solar physics}},\ }\href
  {https://doi.org/10.1007/s10909-012-0574-y} {\bibfield  {journal} {\bibinfo
  {journal} {Journal of Low Temperature Physics}\ }\textbf {\bibinfo {volume}
  {167}},\ \bibinfo {pages} {168} (\bibinfo {year} {2012})}\BibitemShut
  {NoStop}%
\bibitem [{\citenamefont {Smith}\ \emph {et~al.}(2014)\citenamefont {Smith},
  \citenamefont {Adams}, \citenamefont {Bandler}, \citenamefont {Busch},
  \citenamefont {Chervenak}, \citenamefont {Eckart}, \citenamefont
  {Finkbeiner}, \citenamefont {Kelley}, \citenamefont {Kilbourne},
  \citenamefont {Lee}, \citenamefont {Porst}, \citenamefont {Porter},\ and\
  \citenamefont {Sadleir}}]{Smith2014}%
  \BibitemOpen
  \bibfield  {author} {\bibinfo {author} {\bibfnamefont {S.~J.}\ \bibnamefont
  {Smith}}, \bibinfo {author} {\bibfnamefont {J.~S.}\ \bibnamefont {Adams}},
  \bibinfo {author} {\bibfnamefont {S.~R.}\ \bibnamefont {Bandler}}, \bibinfo
  {author} {\bibfnamefont {S.~E.}\ \bibnamefont {Busch}}, \bibinfo {author}
  {\bibfnamefont {J.~A.}\ \bibnamefont {Chervenak}}, \bibinfo {author}
  {\bibfnamefont {M.~E.}\ \bibnamefont {Eckart}}, \bibinfo {author}
  {\bibfnamefont {F.~M.}\ \bibnamefont {Finkbeiner}}, \bibinfo {author}
  {\bibfnamefont {R.~L.}\ \bibnamefont {Kelley}}, \bibinfo {author}
  {\bibfnamefont {C.~A.}\ \bibnamefont {Kilbourne}}, \bibinfo {author}
  {\bibfnamefont {S.~J.}\ \bibnamefont {Lee}}, \bibinfo {author} {\bibfnamefont
  {J.~P.}\ \bibnamefont {Porst}}, \bibinfo {author} {\bibfnamefont {F.~S.}\
  \bibnamefont {Porter}},\ and\ \bibinfo {author} {\bibfnamefont {J.~E.}\
  \bibnamefont {Sadleir}},\ }\bibfield  {title} {\bibinfo {title}
  {{Characterization of Mo/Au transition-edge sensors with different geometric
  configurations}},\ }\href {https://doi.org/10.1007/s10909-013-1031-2}
  {\bibfield  {journal} {\bibinfo  {journal} {Journal of Low Temperature
  Physics}\ }\textbf {\bibinfo {volume} {176}},\ \bibinfo {pages} {356}
  (\bibinfo {year} {2014})}\BibitemShut {NoStop}%
\bibitem [{\citenamefont {Bandler}\ \emph {et~al.}(2008)\citenamefont
  {Bandler}, \citenamefont {Brekosky}, \citenamefont {Brown}, \citenamefont
  {Chervenak}, \citenamefont {Figueroa-Feliciano}, \citenamefont {Finkbeiner},
  \citenamefont {Iyomoto}, \citenamefont {Kelley}, \citenamefont {Kilbourne},
  \citenamefont {Porter}, \citenamefont {Sadleir},\ and\ \citenamefont
  {Smith}}]{Bandler2008}%
  \BibitemOpen
  \bibfield  {author} {\bibinfo {author} {\bibfnamefont {S.~R.}\ \bibnamefont
  {Bandler}}, \bibinfo {author} {\bibfnamefont {R.~P.}\ \bibnamefont
  {Brekosky}}, \bibinfo {author} {\bibfnamefont {A.~D.}\ \bibnamefont {Brown}},
  \bibinfo {author} {\bibfnamefont {J.~A.}\ \bibnamefont {Chervenak}}, \bibinfo
  {author} {\bibfnamefont {E.}~\bibnamefont {Figueroa-Feliciano}}, \bibinfo
  {author} {\bibfnamefont {F.~M.}\ \bibnamefont {Finkbeiner}}, \bibinfo
  {author} {\bibfnamefont {N.}~\bibnamefont {Iyomoto}}, \bibinfo {author}
  {\bibfnamefont {R.~L.}\ \bibnamefont {Kelley}}, \bibinfo {author}
  {\bibfnamefont {C.~A.}\ \bibnamefont {Kilbourne}}, \bibinfo {author}
  {\bibfnamefont {F.~S.}\ \bibnamefont {Porter}}, \bibinfo {author}
  {\bibfnamefont {J.}~\bibnamefont {Sadleir}},\ and\ \bibinfo {author}
  {\bibfnamefont {S.~J.}\ \bibnamefont {Smith}},\ }\bibfield  {title} {\bibinfo
  {title} {{Performance of TES X-ray microcalorimeters with a novel absorber
  design}},\ }\href {https://doi.org/10.1007/s10909-007-9673-6} {\bibfield
  {journal} {\bibinfo  {journal} {Journal of Low Temperature Physics}\ }\textbf
  {\bibinfo {volume} {151}},\ \bibinfo {pages} {400} (\bibinfo {year}
  {2008})}\BibitemShut {NoStop}%
\bibitem [{\citenamefont {Smith}\ \emph {et~al.}(2013)\citenamefont {Smith},
  \citenamefont {Adams}, \citenamefont {Bailey}, \citenamefont {Bandler},
  \citenamefont {Busch}, \citenamefont {Chervenak}, \citenamefont {Eckart},
  \citenamefont {Finkbeiner}, \citenamefont {Kilbourne}, \citenamefont
  {Kelley}, \citenamefont {Lee}, \citenamefont {Porst}, \citenamefont
  {Porter},\ and\ \citenamefont {Sadleir}}]{Smith2013}%
  \BibitemOpen
  \bibfield  {author} {\bibinfo {author} {\bibfnamefont {S.~J.}\ \bibnamefont
  {Smith}}, \bibinfo {author} {\bibfnamefont {J.~S.}\ \bibnamefont {Adams}},
  \bibinfo {author} {\bibfnamefont {C.~N.}\ \bibnamefont {Bailey}}, \bibinfo
  {author} {\bibfnamefont {S.~R.}\ \bibnamefont {Bandler}}, \bibinfo {author}
  {\bibfnamefont {S.~E.}\ \bibnamefont {Busch}}, \bibinfo {author}
  {\bibfnamefont {J.~A.}\ \bibnamefont {Chervenak}}, \bibinfo {author}
  {\bibfnamefont {M.~E.}\ \bibnamefont {Eckart}}, \bibinfo {author}
  {\bibfnamefont {F.~M.}\ \bibnamefont {Finkbeiner}}, \bibinfo {author}
  {\bibfnamefont {C.~A.}\ \bibnamefont {Kilbourne}}, \bibinfo {author}
  {\bibfnamefont {R.~L.}\ \bibnamefont {Kelley}}, \bibinfo {author}
  {\bibfnamefont {S.-J.}\ \bibnamefont {Lee}}, \bibinfo {author} {\bibfnamefont
  {J.-P.}\ \bibnamefont {Porst}}, \bibinfo {author} {\bibfnamefont {F.~S.}\
  \bibnamefont {Porter}},\ and\ \bibinfo {author} {\bibfnamefont {J.~E.}\
  \bibnamefont {Sadleir}},\ }\bibfield  {title} {\bibinfo {title}
  {{Implications of weak-link behavior on the performance of Mo/Au bilayer
  transition-edge sensors}},\ }\href {https://doi.org/10.1063/1.4818917}
  {\bibfield  {journal} {\bibinfo  {journal} {Journal of Applied Physics}\
  }\textbf {\bibinfo {volume} {114}},\ \bibinfo {pages} {074513} (\bibinfo
  {year} {2013})}\BibitemShut {NoStop}%
\bibitem [{\citenamefont {Nagayoshi}\ \emph {et~al.}(2019)\citenamefont
  {Nagayoshi}, \citenamefont {Ridder}, \citenamefont {Bruijn}, \citenamefont
  {Gottardi}, \citenamefont {Taralli}, \citenamefont {Khosropanah},
  \citenamefont {Akamatsu}, \citenamefont {Visser},\ and\ \citenamefont
  {Gao}}]{Nagayoshi2019}%
  \BibitemOpen
  \bibfield  {author} {\bibinfo {author} {\bibfnamefont {K.}~\bibnamefont
  {Nagayoshi}}, \bibinfo {author} {\bibfnamefont {M.~L.}\ \bibnamefont
  {Ridder}}, \bibinfo {author} {\bibfnamefont {M.~P.}\ \bibnamefont {Bruijn}},
  \bibinfo {author} {\bibfnamefont {L.}~\bibnamefont {Gottardi}}, \bibinfo
  {author} {\bibfnamefont {E.}~\bibnamefont {Taralli}}, \bibinfo {author}
  {\bibfnamefont {P.}~\bibnamefont {Khosropanah}}, \bibinfo {author}
  {\bibfnamefont {H.}~\bibnamefont {Akamatsu}}, \bibinfo {author}
  {\bibfnamefont {S.}~\bibnamefont {Visser}},\ and\ \bibinfo {author}
  {\bibfnamefont {J.-R.}\ \bibnamefont {Gao}},\ }\bibfield  {title} {\bibinfo
  {title} {{Development of a Ti/Au TES Microcalorimeter Array as a Backup
  Sensor for the Athena/X-IFU Instrument}},\ }\href
  {https://doi.org/10.1007/s10909-019-02282-8} {\bibfield  {journal} {\bibinfo
  {journal} {Journal of Low Temperature Physics}\ }\textbf {\bibinfo {volume}
  {199}},\ \bibinfo {pages} {943} (\bibinfo {year} {2019})}\BibitemShut
  {NoStop}%
\bibitem [{\citenamefont {Bruijn}\ \emph {et~al.}(2012)\citenamefont {Bruijn},
  \citenamefont {Gottardi}, \citenamefont {den Hartog}, \citenamefont
  {Hoevers}, \citenamefont {Kiviranta}, \citenamefont {de~Korte},\ and\
  \citenamefont {van~der Kuur}}]{Bruijn2012}%
  \BibitemOpen
  \bibfield  {author} {\bibinfo {author} {\bibfnamefont {M.~P.}\ \bibnamefont
  {Bruijn}}, \bibinfo {author} {\bibfnamefont {L.}~\bibnamefont {Gottardi}},
  \bibinfo {author} {\bibfnamefont {R.~H.}\ \bibnamefont {den Hartog}},
  \bibinfo {author} {\bibfnamefont {H.~F.~C.}\ \bibnamefont {Hoevers}},
  \bibinfo {author} {\bibfnamefont {M.}~\bibnamefont {Kiviranta}}, \bibinfo
  {author} {\bibfnamefont {P.~A.~J.}\ \bibnamefont {de~Korte}},\ and\ \bibinfo
  {author} {\bibfnamefont {J.}~\bibnamefont {van~der Kuur}},\ }\bibfield
  {title} {\bibinfo {title} {{High-Q LC Filters for FDM Read out of Cryogenic
  Sensor Arrays}},\ }\href {https://doi.org/10.1007/s10909-011-0422-5}
  {\bibfield  {journal} {\bibinfo  {journal} {Journal of Low Temperature
  Physics}\ }\textbf {\bibinfo {volume} {167}},\ \bibinfo {pages} {695}
  (\bibinfo {year} {2012})}\BibitemShut {NoStop}%
\bibitem [{\citenamefont {Akamatsu}\ \emph {et~al.}(2016)\citenamefont
  {Akamatsu}, \citenamefont {Gottardi}, \citenamefont {van~der Kuur},
  \citenamefont {de~Vries}, \citenamefont {Ravensberg}, \citenamefont {Adams},
  \citenamefont {Bandler}, \citenamefont {Bruijn}, \citenamefont {Chervenak},
  \citenamefont {Kilbourne}, \citenamefont {Kiviranta}, \citenamefont {van~der
  Linden}, \citenamefont {Jackson},\ and\ \citenamefont
  {Smith}}]{Akamatsu2016a}%
  \BibitemOpen
  \bibfield  {author} {\bibinfo {author} {\bibfnamefont {H.}~\bibnamefont
  {Akamatsu}}, \bibinfo {author} {\bibfnamefont {L.}~\bibnamefont {Gottardi}},
  \bibinfo {author} {\bibfnamefont {J.}~\bibnamefont {van~der Kuur}}, \bibinfo
  {author} {\bibfnamefont {C.~P.}\ \bibnamefont {de~Vries}}, \bibinfo {author}
  {\bibfnamefont {K.}~\bibnamefont {Ravensberg}}, \bibinfo {author}
  {\bibfnamefont {J.~S.}\ \bibnamefont {Adams}}, \bibinfo {author}
  {\bibfnamefont {S.~R.}\ \bibnamefont {Bandler}}, \bibinfo {author}
  {\bibfnamefont {M.~P.}\ \bibnamefont {Bruijn}}, \bibinfo {author}
  {\bibfnamefont {J.~A.}\ \bibnamefont {Chervenak}}, \bibinfo {author}
  {\bibfnamefont {C.~A.}\ \bibnamefont {Kilbourne}}, \bibinfo {author}
  {\bibfnamefont {M.}~\bibnamefont {Kiviranta}}, \bibinfo {author}
  {\bibfnamefont {A.~J.}\ \bibnamefont {van~der Linden}}, \bibinfo {author}
  {\bibfnamefont {B.~D.}\ \bibnamefont {Jackson}},\ and\ \bibinfo {author}
  {\bibfnamefont {S.~J.}\ \bibnamefont {Smith}},\ }\bibfield  {title} {\bibinfo
  {title} {{Development of frequency domain multiplexing for the X-ray Integral
  Field unit (X-IFU) on the Athena}},\ }\href
  {https://doi.org/10.1117/12.2232805} {\bibfield  {journal} {\bibinfo
  {journal} {Space Telescopes and Instrumentation 2016: Ultraviolet to Gamma
  Ray}\ }\textbf {\bibinfo {volume} {9905}},\ \bibinfo {pages} {99055S}
  (\bibinfo {year} {2016})}\BibitemShut {NoStop}%
\bibitem [{\citenamefont {Gottardi}\ \emph {et~al.}(2019)\citenamefont
  {Gottardi}, \citenamefont {van Weers}, \citenamefont {Dercksen},
  \citenamefont {Akamatsu}, \citenamefont {Bruijn}, \citenamefont {Gao},
  \citenamefont {Jackson}, \citenamefont {Khosropanah}, \citenamefont {van~der
  Kuur}, \citenamefont {Ravensberg},\ and\ \citenamefont
  {Ridder}}]{Gottardi2019a}%
  \BibitemOpen
  \bibfield  {author} {\bibinfo {author} {\bibfnamefont {L.}~\bibnamefont
  {Gottardi}}, \bibinfo {author} {\bibfnamefont {H.~J.}\ \bibnamefont {van
  Weers}}, \bibinfo {author} {\bibfnamefont {J.}~\bibnamefont {Dercksen}},
  \bibinfo {author} {\bibfnamefont {H.}~\bibnamefont {Akamatsu}}, \bibinfo
  {author} {\bibfnamefont {M.~P.}\ \bibnamefont {Bruijn}}, \bibinfo {author}
  {\bibfnamefont {J.~R.}\ \bibnamefont {Gao}}, \bibinfo {author} {\bibfnamefont
  {B.~D.}\ \bibnamefont {Jackson}}, \bibinfo {author} {\bibfnamefont
  {P.}~\bibnamefont {Khosropanah}}, \bibinfo {author} {\bibfnamefont
  {J.}~\bibnamefont {van~der Kuur}}, \bibinfo {author} {\bibfnamefont
  {K.}~\bibnamefont {Ravensberg}},\ and\ \bibinfo {author} {\bibfnamefont
  {M.}~\bibnamefont {Ridder}},\ }\bibfield  {title} {\bibinfo {title} {{A
  six-degree-of-freedom micro-vibration acoustic isolator for low-temperature
  radiation detectors based on superconducting transition-edge sensors}},\
  }\bibfield  {journal} {\bibinfo  {journal} {Review of Scientific
  Instruments}\ }\textbf {\bibinfo {volume} {90}},\ \href
  {https://doi.org/10.1063/1.5088364} {10.1063/1.5088364} (\bibinfo {year}
  {2019})\BibitemShut {NoStop}%
\bibitem [{\citenamefont {Hoevers}\ \emph {et~al.}(2005)\citenamefont
  {Hoevers}, \citenamefont {Ridder}, \citenamefont {Germeau}, \citenamefont
  {Bruijn}, \citenamefont {{De Korte}},\ and\ \citenamefont
  {Wiegerink}}]{Hoevers2005}%
  \BibitemOpen
  \bibfield  {author} {\bibinfo {author} {\bibfnamefont {H.~F.}\ \bibnamefont
  {Hoevers}}, \bibinfo {author} {\bibfnamefont {M.~L.}\ \bibnamefont {Ridder}},
  \bibinfo {author} {\bibfnamefont {A.}~\bibnamefont {Germeau}}, \bibinfo
  {author} {\bibfnamefont {M.~P.}\ \bibnamefont {Bruijn}}, \bibinfo {author}
  {\bibfnamefont {P.~A.}\ \bibnamefont {{De Korte}}},\ and\ \bibinfo {author}
  {\bibfnamefont {R.~J.}\ \bibnamefont {Wiegerink}},\ }\bibfield  {title}
  {\bibinfo {title} {{Radiative ballistic phonon transport in silicon-nitride
  membranes at low temperatures}},\ }\href {https://doi.org/10.1063/1.1949269}
  {\bibfield  {journal} {\bibinfo  {journal} {Applied Physics Letters}\
  }\textbf {\bibinfo {volume} {86}},\ \bibinfo {pages} {1} (\bibinfo {year}
  {2005})}\BibitemShut {NoStop}%
\bibitem [{\citenamefont {Lindeman}\ \emph {et~al.}(2004)\citenamefont
  {Lindeman}, \citenamefont {Bandler}, \citenamefont {Brekosky}, \citenamefont
  {Chervenak}, \citenamefont {Figueroa-Feliciano}, \citenamefont {Finkbeiner},
  \citenamefont {Li},\ and\ \citenamefont {Kilbourne}}]{Lindeman2004}%
  \BibitemOpen
  \bibfield  {author} {\bibinfo {author} {\bibfnamefont {M.~A.}\ \bibnamefont
  {Lindeman}}, \bibinfo {author} {\bibfnamefont {S.}~\bibnamefont {Bandler}},
  \bibinfo {author} {\bibfnamefont {R.~P.}\ \bibnamefont {Brekosky}}, \bibinfo
  {author} {\bibfnamefont {J.~A.}\ \bibnamefont {Chervenak}}, \bibinfo {author}
  {\bibfnamefont {E.}~\bibnamefont {Figueroa-Feliciano}}, \bibinfo {author}
  {\bibfnamefont {F.~M.}\ \bibnamefont {Finkbeiner}}, \bibinfo {author}
  {\bibfnamefont {M.~J.}\ \bibnamefont {Li}},\ and\ \bibinfo {author}
  {\bibfnamefont {C.~A.}\ \bibnamefont {Kilbourne}},\ }\bibfield  {title}
  {\bibinfo {title} {{Impedance measurements and modeling of a
  transition-edge-sensor calorimeter}},\ }\href
  {https://doi.org/10.1063/1.1711144} {\bibfield  {journal} {\bibinfo
  {journal} {Review of Scientific Instruments}\ }\textbf {\bibinfo {volume}
  {75}},\ \bibinfo {pages} {1283} (\bibinfo {year} {2004})}\BibitemShut
  {NoStop}%
\bibitem [{\citenamefont {Taralli}\ \emph {et~al.}(2019)\citenamefont
  {Taralli}, \citenamefont {Khosropanah}, \citenamefont {Gottardi},
  \citenamefont {Nagayoshi}, \citenamefont {Ridder}, \citenamefont {Bruijn},\
  and\ \citenamefont {Gao}}]{Taralli2019a}%
  \BibitemOpen
  \bibfield  {author} {\bibinfo {author} {\bibfnamefont {E.}~\bibnamefont
  {Taralli}}, \bibinfo {author} {\bibfnamefont {P.}~\bibnamefont
  {Khosropanah}}, \bibinfo {author} {\bibfnamefont {L.}~\bibnamefont
  {Gottardi}}, \bibinfo {author} {\bibfnamefont {K.}~\bibnamefont {Nagayoshi}},
  \bibinfo {author} {\bibfnamefont {M.~L.}\ \bibnamefont {Ridder}}, \bibinfo
  {author} {\bibfnamefont {M.~P.}\ \bibnamefont {Bruijn}},\ and\ \bibinfo
  {author} {\bibfnamefont {J.~R.}\ \bibnamefont {Gao}},\ }\bibfield  {title}
  {\bibinfo {title} {{Complex impedance of TESs under AC bias using FDM readout
  system}},\ }\href {https://doi.org/10.1063/1.5089739} {\bibfield  {journal}
  {\bibinfo  {journal} {AIP Advances}\ }\textbf {\bibinfo {volume} {9}},\
  \bibinfo {pages} {045324} (\bibinfo {year} {2019})}\BibitemShut {NoStop}%
\bibitem [{\citenamefont {Gottardi}\ \emph {et~al.}(2021)\citenamefont
  {Gottardi}, \citenamefont {de~Wit}, \citenamefont {Taralli}, \citenamefont
  {Nagayashi},\ and\ \citenamefont {Kozorezov}}]{Gottardi2021}%
  \BibitemOpen
  \bibfield  {author} {\bibinfo {author} {\bibfnamefont {L.}~\bibnamefont
  {Gottardi}}, \bibinfo {author} {\bibfnamefont {M.}~\bibnamefont {de~Wit}},
  \bibinfo {author} {\bibfnamefont {E.}~\bibnamefont {Taralli}}, \bibinfo
  {author} {\bibfnamefont {K.}~\bibnamefont {Nagayashi}},\ and\ \bibinfo
  {author} {\bibfnamefont {A.}~\bibnamefont {Kozorezov}},\ }\bibfield  {title}
  {\bibinfo {title} {{Voltage Fluctuations in ac Biased Superconducting
  Transition-Edge Sensors}},\ }\href
  {https://doi.org/10.1103/physrevlett.126.217001} {\bibfield  {journal}
  {\bibinfo  {journal} {Physical Review Letters}\ }\textbf {\bibinfo {volume}
  {126}},\ \bibinfo {pages} {217001} (\bibinfo {year} {2021})}\BibitemShut
  {NoStop}%
\bibitem [{\citenamefont {Smith}\ \emph {et~al.}(2015)\citenamefont {Smith},
  \citenamefont {Adams}, \citenamefont {Bandler}, \citenamefont
  {Betancourt-Martinez}, \citenamefont {Chervenak}, \citenamefont {Eckart},
  \citenamefont {Finkbeiner}, \citenamefont {Kelley}, \citenamefont
  {Kilbourne}, \citenamefont {Lee}, \citenamefont {Porter}, \citenamefont
  {Sadleir},\ and\ \citenamefont {Wassell}}]{Smith2015}%
  \BibitemOpen
  \bibfield  {author} {\bibinfo {author} {\bibfnamefont {S.~J.}\ \bibnamefont
  {Smith}}, \bibinfo {author} {\bibfnamefont {J.~S.}\ \bibnamefont {Adams}},
  \bibinfo {author} {\bibfnamefont {S.~R.}\ \bibnamefont {Bandler}}, \bibinfo
  {author} {\bibfnamefont {G.}~\bibnamefont {Betancourt-Martinez}}, \bibinfo
  {author} {\bibfnamefont {J.~A.}\ \bibnamefont {Chervenak}}, \bibinfo {author}
  {\bibfnamefont {M.~E.}\ \bibnamefont {Eckart}}, \bibinfo {author}
  {\bibfnamefont {F.~M.}\ \bibnamefont {Finkbeiner}}, \bibinfo {author}
  {\bibfnamefont {R.~L.}\ \bibnamefont {Kelley}}, \bibinfo {author}
  {\bibfnamefont {C.~A.}\ \bibnamefont {Kilbourne}}, \bibinfo {author}
  {\bibfnamefont {S.~J.}\ \bibnamefont {Lee}}, \bibinfo {author} {\bibfnamefont
  {F.~S.}\ \bibnamefont {Porter}}, \bibinfo {author} {\bibfnamefont {J.~E.}\
  \bibnamefont {Sadleir}},\ and\ \bibinfo {author} {\bibfnamefont {E.~J.}\
  \bibnamefont {Wassell}},\ }\bibfield  {title} {\bibinfo {title} {{Uniformity
  of kilo-pixel arrays of transition-edge sensors for X-ray astronomy}},\
  }\bibfield  {journal} {\bibinfo  {journal} {IEEE Transactions on Applied
  Superconductivity}\ }\textbf {\bibinfo {volume} {25}},\ \href
  {https://doi.org/10.1109/TASC.2014.2369352} {10.1109/TASC.2014.2369352}
  (\bibinfo {year} {2015})\BibitemShut {NoStop}%
\bibitem [{\citenamefont {Kozorezov}\ \emph {et~al.}(2012)\citenamefont
  {Kozorezov}, \citenamefont {Golubov}, \citenamefont {Martin}, \citenamefont
  {de~Korte}, \citenamefont {Lindeman}, \citenamefont {Hijmering},
  \citenamefont {van~der Kuur}, \citenamefont {Hoevers}, \citenamefont
  {Gottardi}, \citenamefont {Kupriyanov},\ and\ \citenamefont
  {Wigmore}}]{Kozorezov2012}%
  \BibitemOpen
  \bibfield  {author} {\bibinfo {author} {\bibfnamefont {A.}~\bibnamefont
  {Kozorezov}}, \bibinfo {author} {\bibfnamefont {A.~A.}\ \bibnamefont
  {Golubov}}, \bibinfo {author} {\bibfnamefont {D.~D.~E.}\ \bibnamefont
  {Martin}}, \bibinfo {author} {\bibfnamefont {P.~A.~J.}\ \bibnamefont
  {de~Korte}}, \bibinfo {author} {\bibfnamefont {M.~A.}\ \bibnamefont
  {Lindeman}}, \bibinfo {author} {\bibfnamefont {R.~A.}\ \bibnamefont
  {Hijmering}}, \bibinfo {author} {\bibfnamefont {J.}~\bibnamefont {van~der
  Kuur}}, \bibinfo {author} {\bibfnamefont {H.~F.~C.}\ \bibnamefont {Hoevers}},
  \bibinfo {author} {\bibfnamefont {L.}~\bibnamefont {Gottardi}}, \bibinfo
  {author} {\bibfnamefont {M.~Y.}\ \bibnamefont {Kupriyanov}},\ and\ \bibinfo
  {author} {\bibfnamefont {J.~K.}\ \bibnamefont {Wigmore}},\ }\bibfield
  {title} {\bibinfo {title} {{Electrical Noise in a TES as a Resistively
  Shunted Conducting Junction}},\ }\href
  {https://doi.org/10.1007/s10909-012-0489-7} {\bibfield  {journal} {\bibinfo
  {journal} {Journal of Low Temperature Physics}\ }\textbf {\bibinfo {volume}
  {167}},\ \bibinfo {pages} {108} (\bibinfo {year} {2012})}\BibitemShut
  {NoStop}%
\bibitem [{\citenamefont {Wessels}\ \emph {et~al.}(2021)\citenamefont
  {Wessels}, \citenamefont {Morgan}, \citenamefont {Gard}, \citenamefont
  {Hilton}, \citenamefont {Mates}, \citenamefont {Reintsema}, \citenamefont
  {Schmidt}, \citenamefont {Swetz}, \citenamefont {Ullom}, \citenamefont
  {Vale},\ and\ \citenamefont {Bennett}}]{Wessels2021}%
  \BibitemOpen
  \bibfield  {author} {\bibinfo {author} {\bibfnamefont {A.}~\bibnamefont
  {Wessels}}, \bibinfo {author} {\bibfnamefont {K.}~\bibnamefont {Morgan}},
  \bibinfo {author} {\bibfnamefont {J.~D.}\ \bibnamefont {Gard}}, \bibinfo
  {author} {\bibfnamefont {G.~C.}\ \bibnamefont {Hilton}}, \bibinfo {author}
  {\bibfnamefont {J.~A.~B.}\ \bibnamefont {Mates}}, \bibinfo {author}
  {\bibfnamefont {C.~D.}\ \bibnamefont {Reintsema}}, \bibinfo {author}
  {\bibfnamefont {D.~R.}\ \bibnamefont {Schmidt}}, \bibinfo {author}
  {\bibfnamefont {D.~S.}\ \bibnamefont {Swetz}}, \bibinfo {author}
  {\bibfnamefont {J.~N.}\ \bibnamefont {Ullom}}, \bibinfo {author}
  {\bibfnamefont {L.~R.}\ \bibnamefont {Vale}},\ and\ \bibinfo {author}
  {\bibfnamefont {D.~A.}\ \bibnamefont {Bennett}},\ }\bibfield  {title}
  {\bibinfo {title} {{A model for excess Johnson noise in superconducting
  transition-edge sensors}},\ }\href {https://doi.org/10.1063/5.0043369}
  {\bibfield  {journal} {\bibinfo  {journal} {Applied Physics Letters}\
  }\textbf {\bibinfo {volume} {118}},\ \bibinfo {pages} {202601} (\bibinfo
  {year} {2021})}\BibitemShut {NoStop}%
\bibitem [{\citenamefont {Wakeham}\ \emph {et~al.}(2019)\citenamefont
  {Wakeham}, \citenamefont {Adams}, \citenamefont {Bandler}, \citenamefont
  {Beaumont}, \citenamefont {Chervenak}, \citenamefont {Datesman},
  \citenamefont {Eckart}, \citenamefont {Finkbeiner}, \citenamefont {Hummatov},
  \citenamefont {Kelley}, \citenamefont {Kilbourne}, \citenamefont {Miniussi},
  \citenamefont {Porter}, \citenamefont {Sadleir}, \citenamefont {Sakai},
  \citenamefont {Smith},\ and\ \citenamefont {Wassell}}]{Wakeham2019}%
  \BibitemOpen
  \bibfield  {author} {\bibinfo {author} {\bibfnamefont {N.~A.}\ \bibnamefont
  {Wakeham}}, \bibinfo {author} {\bibfnamefont {J.~S.}\ \bibnamefont {Adams}},
  \bibinfo {author} {\bibfnamefont {S.~R.}\ \bibnamefont {Bandler}}, \bibinfo
  {author} {\bibfnamefont {S.}~\bibnamefont {Beaumont}}, \bibinfo {author}
  {\bibfnamefont {J.~A.}\ \bibnamefont {Chervenak}}, \bibinfo {author}
  {\bibfnamefont {A.~M.}\ \bibnamefont {Datesman}}, \bibinfo {author}
  {\bibfnamefont {M.~E.}\ \bibnamefont {Eckart}}, \bibinfo {author}
  {\bibfnamefont {F.~M.}\ \bibnamefont {Finkbeiner}}, \bibinfo {author}
  {\bibfnamefont {R.}~\bibnamefont {Hummatov}}, \bibinfo {author}
  {\bibfnamefont {R.~L.}\ \bibnamefont {Kelley}}, \bibinfo {author}
  {\bibfnamefont {C.~A.}\ \bibnamefont {Kilbourne}}, \bibinfo {author}
  {\bibfnamefont {A.~R.}\ \bibnamefont {Miniussi}}, \bibinfo {author}
  {\bibfnamefont {F.~S.}\ \bibnamefont {Porter}}, \bibinfo {author}
  {\bibfnamefont {J.~E.}\ \bibnamefont {Sadleir}}, \bibinfo {author}
  {\bibfnamefont {K.}~\bibnamefont {Sakai}}, \bibinfo {author} {\bibfnamefont
  {S.~J.}\ \bibnamefont {Smith}},\ and\ \bibinfo {author} {\bibfnamefont
  {E.~J.}\ \bibnamefont {Wassell}},\ }\bibfield  {title} {\bibinfo {title}
  {{Thermal fluctuation noise in Mo/Au superconducting transition-edge sensor
  microcalorimeters}},\ }\href {https://doi.org/10.1063/1.5086045} {\bibfield
  {journal} {\bibinfo  {journal} {Journal of Applied Physics}\ }\textbf
  {\bibinfo {volume} {125}},\ \bibinfo {pages} {164503} (\bibinfo {year}
  {2019})}\BibitemShut {NoStop}%
\bibitem [{\citenamefont {Palosaari}\ \emph {et~al.}(2012)\citenamefont
  {Palosaari}, \citenamefont {Kinnunen}, \citenamefont {Maasilta},
  \citenamefont {Ridder}, \citenamefont {van~der Kuur},\ and\ \citenamefont
  {Hoevers}}]{Palosaari2012}%
  \BibitemOpen
  \bibfield  {author} {\bibinfo {author} {\bibfnamefont {M.~R.}\ \bibnamefont
  {Palosaari}}, \bibinfo {author} {\bibfnamefont {K.~M.}\ \bibnamefont
  {Kinnunen}}, \bibinfo {author} {\bibfnamefont {I.~J.}\ \bibnamefont
  {Maasilta}}, \bibinfo {author} {\bibfnamefont {M.}~\bibnamefont {Ridder}},
  \bibinfo {author} {\bibfnamefont {J.}~\bibnamefont {van~der Kuur}},\ and\
  \bibinfo {author} {\bibfnamefont {H.~F.}\ \bibnamefont {Hoevers}},\
  }\bibfield  {title} {\bibinfo {title} {{Analysis of impedance and noise data
  of an x-ray transition-edge sensor using complex thermal models}},\ }\href
  {https://doi.org/10.1007/s10909-012-0471-4} {\bibfield  {journal} {\bibinfo
  {journal} {Journal of Low Temperature Physics}\ }\textbf {\bibinfo {volume}
  {167}},\ \bibinfo {pages} {129} (\bibinfo {year} {2012})}\BibitemShut
  {NoStop}%
\bibitem [{\citenamefont {Maasilta}(2012)}]{Maasilta2012a}%
  \BibitemOpen
  \bibfield  {author} {\bibinfo {author} {\bibfnamefont {I.~J.}\ \bibnamefont
  {Maasilta}},\ }\bibfield  {title} {\bibinfo {title} {{Complex impedance,
  responsivity and noise of transition-edge sensors: Analytical solutions for
  two- and three-block thermal models}},\ }\bibfield  {journal} {\bibinfo
  {journal} {AIP Advances}\ }\textbf {\bibinfo {volume} {2}},\ \href
  {https://doi.org/10.1063/1.4759111} {10.1063/1.4759111} (\bibinfo {year}
  {2012})\BibitemShut {NoStop}%
\bibitem [{\citenamefont {Tinkham}(1996)}]{Tinkham1996}%
  \BibitemOpen
  \bibfield  {author} {\bibinfo {author} {\bibfnamefont {M.}~\bibnamefont
  {Tinkham}},\ }\href@noop {} {\emph {\bibinfo {title} {{Introduction to
  superconductivity}}}},\ \bibinfo {edition} {2nd}\ ed.\ (\bibinfo  {publisher}
  {McGraw-Hill, Inc.},\ \bibinfo {year} {1996})\BibitemShut {NoStop}%
\bibitem [{\citenamefont {Wakeham}\ \emph {et~al.}(2020)\citenamefont
  {Wakeham}, \citenamefont {Adams}, \citenamefont {Bandler}, \citenamefont
  {Beaumont}, \citenamefont {Chang}, \citenamefont {Chervenak}, \citenamefont
  {Eckart}, \citenamefont {Finkbeiner}, \citenamefont {Ha}, \citenamefont
  {Hummatov}, \citenamefont {Kelley}, \citenamefont {Kilbourne}, \citenamefont
  {Miniussi}, \citenamefont {Muramatsu}, \citenamefont {Porter}, \citenamefont
  {Sadleir}, \citenamefont {Sakai}, \citenamefont {Smith},\ and\ \citenamefont
  {Wassell}}]{Wakeham2020}%
  \BibitemOpen
  \bibfield  {author} {\bibinfo {author} {\bibfnamefont {N.~A.}\ \bibnamefont
  {Wakeham}}, \bibinfo {author} {\bibfnamefont {J.}~\bibnamefont {Adams}},
  \bibinfo {author} {\bibfnamefont {S.}~\bibnamefont {Bandler}}, \bibinfo
  {author} {\bibfnamefont {S.}~\bibnamefont {Beaumont}}, \bibinfo {author}
  {\bibfnamefont {M.}~\bibnamefont {Chang}}, \bibinfo {author} {\bibfnamefont
  {J.}~\bibnamefont {Chervenak}}, \bibinfo {author} {\bibfnamefont
  {M.}~\bibnamefont {Eckart}}, \bibinfo {author} {\bibfnamefont
  {F.}~\bibnamefont {Finkbeiner}}, \bibinfo {author} {\bibfnamefont
  {J.}~\bibnamefont {Ha}}, \bibinfo {author} {\bibfnamefont {R.}~\bibnamefont
  {Hummatov}}, \bibinfo {author} {\bibfnamefont {R.~L.}\ \bibnamefont
  {Kelley}}, \bibinfo {author} {\bibfnamefont {C.~A.}\ \bibnamefont
  {Kilbourne}}, \bibinfo {author} {\bibfnamefont {A.~R.}\ \bibnamefont
  {Miniussi}}, \bibinfo {author} {\bibfnamefont {H.}~\bibnamefont {Muramatsu}},
  \bibinfo {author} {\bibfnamefont {F.~S.}\ \bibnamefont {Porter}}, \bibinfo
  {author} {\bibfnamefont {J.~E.}\ \bibnamefont {Sadleir}}, \bibinfo {author}
  {\bibfnamefont {K.}~\bibnamefont {Sakai}}, \bibinfo {author} {\bibfnamefont
  {S.~J.}\ \bibnamefont {Smith}},\ and\ \bibinfo {author} {\bibfnamefont
  {E.~J.}\ \bibnamefont {Wassell}},\ }\bibfield  {title} {\bibinfo {title}
  {{The impact of transition-edge sensor design on internal thermal fluctuation
  noise and thermal conductance}},\ }\href@noop {} {\bibfield  {journal}
  {\bibinfo  {journal} {Poster Presentation ASC}\ } (\bibinfo {year}
  {2020})}\BibitemShut {NoStop}%
\bibitem [{\citenamefont {Gottardi}\ \emph {et~al.}(2018)\citenamefont
  {Gottardi}, \citenamefont {Smith}, \citenamefont {Kozorezov}, \citenamefont
  {Akamatsu}, \citenamefont {van~der Kuur}, \citenamefont {Bandler},
  \citenamefont {Bruijn}, \citenamefont {Chervenak}, \citenamefont {Gao},
  \citenamefont {den Hartog}, \citenamefont {Jackson}, \citenamefont
  {Khosropanah}, \citenamefont {Miniussi}, \citenamefont {Nagayoshi},
  \citenamefont {Ridder}, \citenamefont {Sadleir}, \citenamefont {Sakai},\ and\
  \citenamefont {Wakeham}}]{Gottardi2018a}%
  \BibitemOpen
  \bibfield  {author} {\bibinfo {author} {\bibfnamefont {L.}~\bibnamefont
  {Gottardi}}, \bibinfo {author} {\bibfnamefont {S.~J.}\ \bibnamefont {Smith}},
  \bibinfo {author} {\bibfnamefont {A.}~\bibnamefont {Kozorezov}}, \bibinfo
  {author} {\bibfnamefont {H.}~\bibnamefont {Akamatsu}}, \bibinfo {author}
  {\bibfnamefont {J.}~\bibnamefont {van~der Kuur}}, \bibinfo {author}
  {\bibfnamefont {S.~R.}\ \bibnamefont {Bandler}}, \bibinfo {author}
  {\bibfnamefont {M.~P.}\ \bibnamefont {Bruijn}}, \bibinfo {author}
  {\bibfnamefont {J.~A.}\ \bibnamefont {Chervenak}}, \bibinfo {author}
  {\bibfnamefont {J.~R.}\ \bibnamefont {Gao}}, \bibinfo {author} {\bibfnamefont
  {R.~H.}\ \bibnamefont {den Hartog}}, \bibinfo {author} {\bibfnamefont
  {B.~D.}\ \bibnamefont {Jackson}}, \bibinfo {author} {\bibfnamefont
  {P.}~\bibnamefont {Khosropanah}}, \bibinfo {author} {\bibfnamefont {A.~R.}\
  \bibnamefont {Miniussi}}, \bibinfo {author} {\bibfnamefont {K.}~\bibnamefont
  {Nagayoshi}}, \bibinfo {author} {\bibfnamefont {M.}~\bibnamefont {Ridder}},
  \bibinfo {author} {\bibfnamefont {J.~E.}\ \bibnamefont {Sadleir}}, \bibinfo
  {author} {\bibfnamefont {K.}~\bibnamefont {Sakai}},\ and\ \bibinfo {author}
  {\bibfnamefont {N.}~\bibnamefont {Wakeham}},\ }\bibfield  {title} {\bibinfo
  {title} {{Josephson Effects in Frequency-Domain Multiplexed TES
  Microcalorimeters and Bolometers}},\ }\href
  {https://doi.org/10.1007/s10909-018-2006-0} {\bibfield  {journal} {\bibinfo
  {journal} {Journal of Low Temperature Physics}\ }\textbf {\bibinfo {volume}
  {193}},\ \bibinfo {pages} {209} (\bibinfo {year} {2018})}\BibitemShut
  {NoStop}%
\end{thebibliography}%

\end{document}